\documentclass[a4,11pt]{article}
 \usepackage{graphicx}
 \usepackage{amsmath}
 \usepackage{wrapfig}
 \usepackage{rotating}
 \usepackage{lscape}
 \usepackage{longtable}
 \usepackage{dcolumn}
 \usepackage{multirow}
 \usepackage[super,comma,numbers,sort&compress]{natbib}
 \usepackage{ccaption}
 \usepackage{hyperref}
 \usepackage{slashbox}
 \usepackage{subcaption}

  \hypersetup{linkbordercolor={0 0 1}, citebordercolor={0 0 1},
 	colorlinks=true, linkcolor=blue, citecolor=blue}

 \renewcommand{\thetable}{\Roman{table}}

 \setcitestyle{super}
 
 \newcommand*{\citen}[1]{%
 	\begingroup
 	\romannumeral-`\x 
 	\setcitestyle{numbers}%
 	\cite{#1}%
 	\endgroup
 }

\usepackage[pagewise]{lineno}

\begin{document}

\large
\title{Construction of First Principle Based  Adiabatic and Diabatic Hamiltonian for TiO$_6^{8-}$ unit of BaTiO$_3$ Crystal: Photoemission Spectra and Ferroelectricity}
\author{Mantu Kumar Sah{$^a$}, Soumya Mukherjee{$^{a, b}$}, Satyam Ravi{$^c$} and Satrajit Adhikari$^{a,}$\footnote{Author to whom correspondence should be addressed: 
		e-mail: pcsa@iacs.res.in}\\
{$^a$}School of Chemical Sciences, \\ 
Indian Association for the Cultivation of Science, 
Jadavpur, Kolkata - 700032, India\\
{$^{b}$}Department of Chemistry, MANIT, Bhopal, India\\
{$^c$}School of Advanced Sciences, VIT Bhopal University, Bhopal, India}
\date{\today}
\maketitle

\begin{abstract}  
\begin{small} 
\noindent
The ferroelectric property of BaTiO$_3$ crystal arises from the strong Pseudo Jahn-Teller (PJT) interactions between the non-degenerate ground electronic state, $^1A_{1g}$ and the degenerate $^1T_{1u}$ symmetry states through the nuclear distortions of $t_{1u}$ modes in TiO$_6^{8-}$ unit. In a $d^0$ electronic configuration of $Ti^{4+}$ ion, the PJT interaction leads to a stabilization effect, which has been explored using Beyond Born-Oppenheimer (BBO) theory. The $^1T_{1u}$ excited states form a three-state degeneracy, exhibiting feeble Jahn-Teller (JT) distortions over the $t_{2g}$ planes. For the first time,  we compute \textit{ab initio} adiabatic potential energy surfaces (PESs) and non-adiabatic coupling terms (NACTs), and thereafter, diabatic PESs and couplings for the perovskite unit, TiO$_6^{8-}$. Using a Time-Dependent Discrete Variable Representation (TDDVR) approach, the theoretical photoemission spectra exhibit good agreement with the experimental ones.  Moreover, the experimental observation on order parameter associated with  ferroelectric properties of BaTiO$_3$ crystal show close resemblance with present and other theoretical predictions. 
\end{small}
\end{abstract}

\newpage
\section{Introduction}\label{sec:intro}
\vspace{0.2 cm}
\noindent
Over the last few decades, the electronic structure and various spectral properties of TiO$_6^{8-}$ unit of BaTiO$_3$ crystal raised immense interest to the allied scientific community due to its prominent ferroelectric properties originated from strong Pseudo Jahn-Teller (PJT) interactions between the ground electronic state ($^1A_{1g}$) and the excited ones ($^1T_{1u}$). In other words, the profound non-adiabatic couplings among the electronic states ($^1A_{1g}$ and $^1T_{1u}$) play an important role for depicting the ferroelectric properties and complex spectral features. In the late sixties, I. B. Bersuker \cite{Bersuker_1966} first depicted that the adiabatic potential energy surfaces (PESs) of TiO$_6^{8-}$ unit contains eight (8) minima over the planes constituted with $t_{1u}$ symmetric normal modes corresponding to eight distorted nuclear configurations. At $T=0\ K$, the TiO$_6^{8-}$ unit gets trapped in any one of those minima, whereas at higher temperature, that moiety gains sufficient energy to overcome the energy barrier between two minima and therefore, the averaging of all the eight structures associated with those minima leads to an undistorted octahedral nuclear configuration. Such nuclear displacements destroy the inversion symmetry of the octahedral unit leading to temperature dependent phases and spontaneous crystal polarisation, namely, ferroelectricity. \\

\noindent
The TiO$_6^{8-}$ unit of BaTiO$_3$ crystal remains a long standing interest for theoreticians as well as experimentalist due to its PJT active $^1A_{1g}$ symmetric ground state (originated from $d^0$ electronic configuration) interacting with triply degenerate excited one, $^1T_{1u}$. Kole\.{z}y\'{n}ski and co-worker~\cite{Kolezynski_2005} calculated the band structure of BaTiO$_3$ crystal using FP LAPW approach as implemented in WIEN2k program, which depicts broad dispersion of the valence band (about 5 eV) and a narrow forbidden energy gap (about 2 eV). Such calculations were carried out for cubic and tetragonal structures of the crystal so that the changes in the electronic structure due to the ferroelectric phase transition could be investigated. Moreover, their work explored the role of $\pi$ bonds of oxygen atoms in the ferroelectric behaviour of the material. On the other hand, Qi-Jun Liu and co-workers~\cite{Qi-Jun_Liu_2013} performed density functional theory (DFT) calculations to determine the structural, electronic and optical properties of various phases as well as molecular-orbital bonding of the titled system, where the calculated ground state energies, relative stability and structural parameters show good agreement with the experimental results. Later on, Polinger {\it et al.}~\cite{bersuker_pbcm457} studied the origin of dipolar distortion of the titled system using Green's function method augmented by DFT based computations, where it was depicted that those ferroelectric distortions are mainly originated from PJT effects between ground and excited electronic states of opposite parity. In addition, their work unveiled that multi-center long-range dipole-dipole interactions and their magnitudes are governed by the PJT effects. In the subsequent works, I. B. Bersuker predicted~\cite{Bersuker_2015} that the local distortions due to PJT interactions are dynamic in nature, but there would be a static dipole moment resulting into polarization and enhanced flexoelectricity. All the theoretical investigations considering several properties of PJT active perovskite systems, namely, ferroelectricity, multiferroicity, permittivity and flexoelectricity are summarized in the recently published review articles~\cite{Bersuker_review_2021}. On the other hand, to understand the origin of ferroelectric instability in perovskite materials, Cohen \textit{et al.}~\cite{Cohen_ferroelectrics_1992,Cohen_nature_1992} performed comprehensive self-consistent total energy calculations using the linearized augmented plane wave (LAPW) method within the local density approximation (LDA).  Their findings on BaTiO\(_3\) revealed that the hybridization of Ti 3d and O 2p states is crucial for the $Ti-O$ interaction, which drives ferroelectric instability. Furthermore, Junquera and Ghosez~\cite{Junquera_nature_2003} showed that ferroelectric instability of thin films of BaTiO\(_3\)  requires a minimum thickness of about 6 unit cells (about 24 \AA ), below which off-centre instability does not occur. Later on, Garc\'{\i}a-Lastra \textit{et al.}~\cite{Garcia_Inorg_chem_2014} investigated both local and cooperative effects on ferroelectricity in BaTiO\(_3\) using DFT-based LDA calculations. Such an investigation explored the external condition (like pressure) for the movement of single $Ti^{4+}$ ion from the centre depicting the cooperative effects on the modulation of ferroelectricity.\\
 
\noindent
In the last two decades, few theoretical investigations were carried out on the effects of spin-orbit (SO) coupling in metal and metal oxide doped BaTiO$_3$ crystals. Mirhosseini \textit{et. al}~\cite{Mirhosseini_phys_rev_2010} attempted to explore the presence of Rashba spin-orbit coupling in the 6p states of a Bi adlayer on BaTiO$_3$ (001), where electric polarization originated from the ferroelectricity of BaTiO$_3$ is mainly responsible for such SO couplings. Later on, Zhong and coworkers~\cite{Zhong_Adv_Mat} extended such approach for Ba(Os,Ir,Ru)O$_3$ doped BaTiO$_3$ crystals, where giant Rashba SO splittings were unveiled and even monitored by an external electric field. Moreover, structural, electronic, magnetic and mechanical properties of the mixed perovskites (BiFeO$_3$ and BaTiO$_3$) were investigated using DFT based approaches, which depict strong SO effects in BiFeO$_3$ due to BaTiO$_3$ crystal.~\cite{Ali_BFO_BTO}  On the other hand, theoretical study of SO coupling in pure BaTiO$_3$ crystal is yet an unexplored area.\\

\noindent
 Several experimental
 groups~\cite{Pinczuki_1967,Turik_Shevchenko_1979,R_Cord_1985,Roberts_1987,Nasser_Expt_2000,Wan_2008} were also involved to reveal the unique properties of perovskite systems like TiO$_6^{8-}$ 
with $d^0$ electronic configuration. In the late sixties, Pinczuk {\it et al.} \cite{Pinczuki_1967} determined the symmetries and frequencies of various Raman active vibrational modes of BaTiO$_3$ crystal using various scattering geometries. Turik and Shevchenko~\cite{Turik_Shevchenko_1979} measured its dielectric spectra over a wide range of frequencies (10$^3$ to 10$^{10}$ Hz), where the dielectric permittivity ($\epsilon_{33}$) of the single-domain crystal undergoes a sharp jump around 10$^7$ Hz indicating the phase transition of BaTiO$_3$ crystal. The ultraviolet photoemission spectroscopy (UPS) of the valence electronic level, with a photon energy ($h\nu$) of 105 eV, was conducted on sputtered and annealed single crystal of BaTiO$_3$ by B. Cord and R. Courths.~\cite{R_Cord_1985}  Other groups~\cite{Roberts_1987,Nasser_Expt_2000} have performed X-ray diffraction studies as well as recorded X-ray photoelectron spectra of the BaTiO$_3$ thin films, which contain ions of various oxidation states of titanium atoms, namely, $Ti^{2+}$, $Ti^{3+}$ and $Ti^{4+}$. In a recent work, Wan {\it et al.}~\cite{Wan_2008} investigated the low frequency dielectric and optical properties of powdered BaTiO$_3$ using terahertz time-domain spectroscopy (THz-TDS), which clearly indicates that the dielectric constant of BaTiO$_3$ has a strong correlation with the lowest optical phonon modes at 180 cm$^{-1}$. Such experimental study establishes a relation between low-frequency dielectric function with the optical phonon mode for ferroelectric materials.\\ 

\noindent
The theoretical investigation~\cite{Kolezynski_2005,Qi-Jun_Liu_2013,bersuker_pbcm457,Bersuker_2015,Bersuker_review_2021} on TiO$_6^{8-}$ unit of BaTiO$_3$ crystal predicted the origin of ferroelectric properties due to the PJT interaction among the electronic states ($^1A_{1g}$ and $^1T_{1u}$) through the $t_{1u}$ modes, where the experimental studies~\cite{Pinczuki_1967,Turik_Shevchenko_1979,R_Cord_1985,Roberts_1987,Nasser_Expt_2000,Wan_2008} on Raman active modes, dielectric optical properties and in particular, terahertz spectroscopy of powdered BaTiO$_3$ depicted strong electron-nuclear correlations  leading to such ferroelectric material. In this context, we perform {\it ab initio} (MRCI) calculation to generate adiabatic potential energy surfaces (PESs) and non-adiabatic coupling terms (NACTs) of TiO$_6^{8-}$ unit assuming the crystal structure of BaTiO$_3$, where eight TiO$_6^{8-}$ unit share one Barium (Ba) atom in an octahedral environment. Since the lowest four electronic states ($^1A_{1g}$ and $^1T_{1u}$) form the sub-Hilbert space~\cite{mantu_curl} over the nuclear planes constituted with the normal modes ($t_{1u}$ and $t_{2g}$), we employ Beyond Born-Oppenheimer (BBO) theory and construct the diabatic Hamiltonian for the system. Indeed, it is challenging issue due to the presence of $d^0$ transition metals, like $Ti^{4+}$ possessing three state degeneracy, which would be the first time in literature.\\

\noindent
The theoretical framework of Born-Oppenheimer (BO) treatment~\cite{BO} followed by Born and Huang~\cite{BH} approach  can be considered as the starting point to study multi-mode multi-surface molecular processes and chemical reactions. The two physically meaningful quantities, namely adiabatic potential energy surfaces (PESs) and non-adiabatic coupling terms (NACTs) are the major outcomes of the BO treatment, where both of them dictate the nuclear motion in physical-chemical processes. In other words, the eigenvalues of electronic Hamiltonian are known as adiabatic PESs on which the nuclei move and such movements of nuclei from one electronic state to another are due to the electron-nuclear couplings, also defined  as the momentum couplings (NACTs) between different electronic states. On the contrary, when the interactions among the electronic states are neglected by imposing BO approximation for low energy processes, the dynamics of nuclei are presumably restricted on specific electronic states, but even for ground state, significant anomalies are observed between theoretically calculated and experimentally measured spectral profiles and/or scattering cross-sections obtained from multi-state molecular process or chemical reactions.~\citep{last_jcp107,baer_jcp105,adhikari_jcp111,varandas_jcp112,adhikari_pra62,sarkar_jcp124,sarkar_jpca112,akpaul_jcp131} \\  

\noindent
Since the electron-nuclear couplings (NACTs) exhibit singular behaviour at and around the degenerate points (CIs) over the nuclear configuration space (CS) as depicted by Hellmann-Feynman theorem~\citep{hellmann,feynmann}, a paradigm shift is required by transforming the adiabatic representation of Schr\"odinger equation (SE) to the diabatic framework, where those singular non-adiabatic (kinetic) couplings turn into well-behaved and smooth diabatic (potential) couplings. In the realm of diabatization schemes for realistic systems, a wide variety of approaches were introduced, namely, vibronic coupling model,~\citep{kdc_acp57,koppel_jcp110,koppel_jcp150,viel_jcp124} exact factorisation scheme,~\citep{gross_prl105,gross_jcp137} direct dynamics approaches~\citep{worth_irpc27,worth_irpc34} and many others~\citep{varandas_jcp112,varandas_jcp86,takatsuka_rev,takatsuka_pccp13_2011,TODD:ARPC,martinez_chemrev}. Nevertheless, the first principle (BBO) based adiabatic-to-diabatic transformation (ADT) formalism by F. T. Smith~\citep{smith_pr179} for diatomic species followed by its non-trivial generalization by 
M. Baer~\citep{baer_jcp66,baer_cpl35,baer_phys_rep358,baer_book,baer_jcp117} for triatomic chemical reactions as well as polyatomic molecular processes can be considered as the most ``accurate" methodology for constructing diabatic Hamiltonian for a molecular system/chemical process. In the Baer's formalism~\citep{baer_jcp66,baer_cpl35,baer_phys_rep358}, one needs to integrate a set of coupled differential (ADT) equations along two-dimensional contours over the interested domain of nuclear CS for a given sub-Hilbert space (SHS). If the initially chosen SHS forms a ``true" sub-space, the ``accuracy" of the diabatic PESs and couplings can be affirmed by exploring the validity of two important conditions, namely, (a) the mixing (ADT) angles along a loop enclosing $n$ number of JT type CI(s) or RT interaction(s) or PJT coupling attain the  integer ($n$) multiple of $\pi$ or $2\pi$ or zero (0), respectively~\citep{baer_phys_rep358,baer_mp75,baer_jpca104}; and (b) the vector fields associated with the NACTs obey ``Curl Condition"~\citep{baer_phys_rep358,baer_book,mantu_curl}. On the other hand, later on, Adhikari and coworkers~\citep{sarkar_jcp124,sarkar_jpca112,soumya_jpca121,soumya_jcp150,naskar_jctc}
have developed the first principle based BBO treatment for three to six state SHSs with an explicit formulation of NACTs as well as various ADT quantities (ADT equations, Curl-divergence equations and diabatic PESs matrix elements), which are being widely used for several realistic spectroscopically interesting
molecules~\citep{mukherjee_jpca117,mukherjee_cp440,sardar_cp416,mukherjee_jcp143,soumya_jpca121,bijit_mp115,soumya_acs,bijit_cp515,soumya_jcp150,soumya_comptc,bijit_jpcs1148,bijit_irpc38,joy_pccp,soumya_bz,soumya_tfb,saikat_pyra,Mantu_2023,Mukherjee_2024}, solid state perovskites~\citep{joy_jpca} as well as triatomic scattering processes~\citep{mukherjee_jcp141,bijit_jpcs759,bijit_jpcs833,bijit_jpcs1148,jcp-147-074105-SA,bijit_fh2,bijit_irpc38,joy_pccp} to depict the workability of those ADT equations. In our recent works, we have generalized the BBO formalism by developing a user-friendly OpenMP parallelized program package, `ADT'~\citep{naskar_jctc} for symbolic formulation as well as numerical calculation of the ADT quantities (ADT angles, ADT matrices and diabatic PESs matrices) for any molecular system/chemical processes with $N$ coupled electronic states and $K$ number of nuclear degrees of freedom (DOFs). \\

\noindent
The present article is organized as follows: Section \ref{Theoretical Backgroundy} depicts the theoretical background of BBO theory as well as the working equations to compute the PE spectra and to explore the origin of ferroelectricity. On the other hand, Sections \ref{sec:abinitio} and \ref{sec:adt} demonstrate the details of {\it ab initio} calculations and functional forms of various {\it ab initio} quantities (adiabatic PESs and NACTs), respectively. The numerical results of various ADT quantities (ADT angles, ADT matrices and diabatic PESs matrices) are presented in Subsections \ref{ssec:pair1} and \ref{ssec:pair2}. The TDDVR calculated PE spectra are depicted in Subsections \ref{Photoemission Spectra} and the parametric dependence on ferroelectric properties are presented in Subsection \ref{Ferroelectricity}. Finally, the conclusion of the entire work is laid down in Section \ref{Conclusion}.

\section{Theoretical Background} \label{Theoretical Backgroundy}
\subsection{Brief Outline of Beyond Born-Oppenheimer Theory}
\noindent
Since the molecular processes and chemical reactions are governed by the total electron-nuclear Hamiltonian, the associated 
molecular Schr\"odinger Equation (SE) takes the following form in terms of total Hamiltonian ($\hat{H}$) and wavefunction ($\Psi$):
\begin{eqnarray}
\hat{H}(\textbf{r$_e$},\textbf{R}_n)\Psi(\textbf{r$_e$},\textbf{R}_n) = E\Psi(\textbf{r$_e$},\textbf{R}_n),
\label{mse}
\end{eqnarray}
where \textbf{r$_e$} and \textbf{R$_n$} signify electronic and nuclear coordinate vectors, respectively. For non-relativistic
case, molecular Hamiltonian $\hat{H}(\textbf{r}_e,\textbf{R}_n)$ can be segregated into nuclear kinetic energy operator 
($\hat{T}_{\rm nuc}(\textbf{R}_n)$) and electronic Hamiltonian ($\hat{H}_{\rm{el}}(\textbf{r}_e;\textbf{R}_n)$) as follows:
\begin{eqnarray}
\hat{H}(\textbf{r$_e$},\textbf{R}_n) = \hat{T}_{\rm nuc}(\textbf{R}_n) + \hat{H}_{\rm{el}}(\textbf{r$_e$};\textbf{R}_n)
; \quad \hat{T}_{\rm nuc} = -\frac{\hbar^2}{2}\nabla^2_{R},
\label{eq:hpart}
\end{eqnarray} 

\noindent
where $\boldsymbol{\nabla_{R}}$ symbolizes derivatives on mass-weighted nuclear coordinates (\textbf{R$_n$}). \\

\noindent
For a complete matrix (Hilbert) space of dimension $M$, total molecular wavefunction [$\Psi(\textbf{r}_e,\textbf{R}_n)$] 
can be written as linear combination of electronic eigenfunctions (\{$\xi_i(\textbf{r}_e;\textbf{R}_n)$\}s) through Born-Oppenheimer-Huang~\citep{BO,BH} expansion: 
\begin{equation}
\Psi(\textbf{r}_e,\textbf{R}_n) = \sum_{i=1}^M \psi_i^{\rm{ad}}(\textbf{R}_n) \xi_i(\textbf{r}_e;\textbf{R}_n), 
\label{eq:bo-expansion} 
\end{equation}

\noindent
where the combining coefficients are denoted as nuclear wavefunctions, \{$\psi_i^{\rm{ad}}(\textbf{R}_n)$\}. On the other hand, the electronic eigenfunctions (\{$\xi_i(\textbf{r}_e;\textbf{R}_n)$\}s) are originated from the electronic SE:
\begin{equation}
\hat{H}_{\rm{el}}(\textbf{r}_e;\textbf{R}_n)\xi_i(\textbf{r}_e;\textbf{R}_n) = u_i(\textbf{R}_n)
\xi_i(\textbf{r}_e;\textbf{R}_n), \qquad  
\langle \xi_i | \xi_j \rangle_{\textbf{r}_e} = \delta_{ij},
\label{eq:elese}
\end{equation}
where the eigenvalues, namely, adiabatic PESs (\{$u_i(\textbf{R}_n)$\}s) depend on nuclear coordinates. In 
Eq. \ref{eq:elese}, the expression $\langle \cdots \rangle_{\textbf{r}_e}$ represents the inner product between two electronic eigenstates over electronic 
coordinates. \\

\noindent
While substituting the BO expanded molecular wavefunction (Eq. \ref{eq:bo-expansion}), total Hamiltonian (Eq. \ref{eq:hpart}) 
and electronic SE (Eq. \ref{eq:elese}) in the molecular SE (Eq. \ref{mse}) and then, projecting with various electronic eigenfunctions,
the kinetically coupled nuclear SE (known as adiabatic nuclear SE) takes the following form,
\begin{equation}
-\frac{\hbar^2}{2}\left(\boldsymbol{\nabla}_{R}+\boldsymbol{\tau}\right)^2\psi^{\rm{ad}} + \left(U-E\right)
\psi^{\rm{ad}} = 0,
\label{eq:adsem}
\end{equation}
where the elements of diagonal adiabatic PESs matrix and skew-symmetric non-adiabatic coupling matrix [$\boldsymbol{\tau}_{ij}(\textbf{R}_n) (= \langle\xi_i\vert\boldsymbol{\nabla}_{R}\xi_j\rangle_{\textbf{r}_e})$] can be written as:
\begin{equation}
U_{ij}(\textbf{R}_n)\,=u_j(\textbf{R}_n)\delta_{ij}
\label{eq:adiael}
\end{equation} 
\begin{equation}
\langle 
\xi_j | \boldsymbol{\nabla}_{R}\xi_i\rangle_{\textbf{r}_e} = -\langle 
\xi_i | \boldsymbol{\nabla}_{R}\xi_j\rangle_{\textbf{r}_e}
\label{eq:skew}
\end{equation} 

\vspace{0.5 cm}

\noindent
If non-adiabatic matrix ($\tau$) is dropped from Eq.~\ref{eq:adsem}, the coupling between electronic states will be absent, which is known as Born-Oppenheimer approximation. On the contrary, employing Hellmann-Feynman theorem,~\cite{hellmann,feynmann} the electronic SE (Eq. \ref{eq:elese}) depicts that NACTs (see Eq. \ref{eq:skew}) encounter singularity at the close neighbourhood of degenerate point(s) over the nuclear configuration space (CS):     
\begin{eqnarray}
\boldsymbol{\tau}_{ij} = \dfrac{\langle\xi_i\vert\boldsymbol{\nabla}_{R} 
\hat{H}_{\rm{el}} \vert \xi_j\rangle_{\textbf{r}_e}}{u_j-u_i}\,\, .
\label{eq:hf}
\end{eqnarray}

\noindent
Therefore, a paradigm shift from adiabatic to diabatic representation of nuclear SE is indeed necessary to wipe out the often singular kinetic (NACTs) couplings by well-behaved potential (diabatic) couplings to avoid the 
numerical inaccuracies that could arise in dynamical calculations with adiabatic SE. Such changeover (adiabatic to diabatic) is carried out using the transformation $\psi^{\rm ad}=A\psi^{\rm dia}$, where $\psi^{\rm ad}$ and $\psi^{\rm dia}$ represent adiabatic and diabatic basis, respectively. After substituting such relationship ($\psi^{\rm ad}=A\psi^{\rm dia}$) in Eq. \ref{eq:adsem}, one can achieve the following diabatic representation of nuclear SE:
\begin{eqnarray}
\left(-\dfrac{\hbar^2}{2}\nabla_{R}^2 + W-E\right)\psi^{\rm{dia}} = 0, \quad W = A^{\dagger}U A,
\label{eq:diaben}
\end{eqnarray}

\noindent
under the Adiabatic-to-Diabatic Transformation (ADT) condition:~\citep{baer_phys_rep358} 
\begin{eqnarray}
\boldsymbol{\nabla}_{R}A + \boldsymbol{\tau}A = 0.
\label{eq:adteq}
\end{eqnarray}
\noindent
Since $\boldsymbol{\tau}$ matrix for $M$ dimensional Hilbert space is skew-symmetric in nature, the above-mentioned ADT condition affirms 
the orthogonal property of $A$ matrix [$SO(M)$]. The generalized form of $A$ matrix is depicted elsewhere~\citep{bijit_irpc38,saikat_pyra}.

\vspace{0.2 cm}

\noindent
While substituting the generalized form of ADT ($A$) and the skew-symmetric $\boldsymbol{\tau}$ matrix in ADT condition (Eq. \ref{eq:adteq}),
we obtain $^NC_2\,(=P)$ number of first order differential equations as follows:  
\begin{eqnarray}
\boldsymbol{\nabla}_R\Theta_{ij} = \sum_{m=1}^{P} c^{(m)}\boldsymbol{\tau}^{(m)},
\label{eq:ADTgen}
\end{eqnarray}

\noindent
where $\boldsymbol{\tau}^{(m)}$ signify NACTs between different electronic states and the associated coefficients, $\{c^{(m)}\}$s appear as trigonometric functions of ADT angles $\{\Theta_{kl}\}$ (see Section S3 of the Electronic Supplementary Information (ESI)). Employing a package program~\cite{naskar_jctc}, the details of such formulation of differential equation for any chemical or physical process involving $N$ electronic states and $M$ vibrational modes can be obtained through symbolic manipulation, where the numerical solution of ADT angles ($\{\Theta_{ij}\}$) using \textit{ab initio} calculated NACTs could be carried out.

\subsubsection{Conditions for the Existence of Sub-Hilbert Space}

\noindent
While transforming the adiabatic SE (Eq.~\ref{eq:adsem}) to Diabatic one (Eq.~\ref{eq:diaben}), it is necessary  to satisfied ADT equation (see Eq.~\ref{eq:adteq}). Since ADT is a set of differential equation involved with vector fields, $\boldsymbol\tau$ and those fields could often be singular (see Eq. \ref{eq:hf}) in the nuclear CS, such molecular fields have to keep the conservation of energy for a chosen SHS associated with a molecular process or a chemical reaction. In other words, with the cross derivatives on the scalar components of ADT equation lead to the ADT condition that has one to one correspondence with the curl condition of a vector field as appears in classical electrodynamics  (for more details, see Section S2 of the ESI). If such condition is satisfied for a chemical or physical process involving certain number of electronic states, we may safely predict that those states form a SHS over interested energy domain. Moreover, as the NACTs ($\boldsymbol\tau$) could be singular (see Eq.~\ref{eq:hf}), those terms will depict an integer multiple of $\pi$ on contour integration like any other singular function, if the \textit{ab initio} NACTs ($\boldsymbol\tau$) are calculated  with appropriately defined sub-Hilbert space.

\vspace{0.2 cm}

\noindent
For a sub-Hilbert space, the NACT components should obey the following conditions for two-dimensional ($p$-$q$) cross-section  of $M$ dimensional nuclear degrees of freedom (DOFs):\\

\noindent
(a) Cauchy's residue theorem: Quantization of NACTs over the polar counterpart, $\rho-\phi$:
\begin{equation}
\int_0^{2\pi} \tau_{ij}^{\phi}(\rho_0,\phi)d\phi = 0 \,\,\text{or}\,\,n\pi\,\,\text{or}\,\,n\cdot 2\pi,
\label{eq:quant}
\end{equation}
\noindent
validates the presence of PJT interactions or $n$ number of JT or RT interactions, respectively, encapsulated by the closed contour of integration, where $\tau_{ij}^{\phi}(\rho_0,\phi)$ represents angular ($\phi$) component of NACT between $\lbrace i,j\rbrace$ electronic states along the angular coordinate $\phi$ at $\rho=\rho_0$. \\
  
\noindent
 (b) When we take cross derivatives on the scalar components of ADT conditions (Eq. \ref{eq:adteq}), the Curl conditions appear as:
\begin{equation}
\dfrac{\partial}{\partial p}\tau_{ij}^q-\dfrac{\partial}{\partial q}\tau_{ij}^p=(\tau^q\tau^p)_{ij}-
(\tau^p\tau^q)_{ij},
\label{eq:curl}
\end{equation}
\noindent
which needs to be fulfilled by the $p$ and $q$ components of NACTs and the associated matrix elements
\begin{equation}
F^{pq}_{ij}=\Big[\dfrac{\partial}{\partial p}\tau_{ij}^q-\dfrac{\partial}{\partial q}\tau_{ij}^p\Big]-\Big[(\tau^q\tau^p)_{ij}-
(\tau^p\tau^q)_{ij}\Big]=Z^{pq}_{ij}-C^{pq}_{ij},
\label{eq:curl1}
\end{equation}
\noindent
should acquire the magnitude of zero (0).

\vspace{0.1 cm}

\noindent
Even though, in principle, the above two conditions are valid for a complete Hilbert space, it is important to explore the existence of a sub-Hilbert space for all ``practical" purposes (of computation) and to construct finite [$N\,(<< M)$] dimensional ``accurate"  diabatic Hamiltonian matrix. In other words, for an initially chosen sub-Hilbert space (SHS), if the above conditions (Eqs. \ref{eq:quant} and \ref{eq:curl1}) are not fulfilled, the sub-space needs to be expanded to achieve desired level of accuracy in the diabatic Hamiltonian so that the non-removable component of NACTs becomes negligible and the removable part can be ``truly" diabatized. 

\subsubsection{Path Dependence of ADT vis-\'a-vis Diabatic PES Matrices}\label{sec:path}

\noindent
Since the BBO theory involves skew-symmetric non-Abelian non-adiabatic coupling matrix (NACM) for three or higher dimensional sub-Hilbert space, the solution of ADT (coupled differential) equations can be path dependent with the choice of contour over the two-dimensional cross-sections of multi-dimensional space of nuclear degrees of freedom. For $p$-$q$ nuclear plane, if the integral forms of the orthogonal ADT matrices for two different paths with same final nuclear configuration are represented by $A_1$ and $A_2$~\citep{mukherjee_jpca117}, the associated diabatic PESs matrices ending at same point can be represented as: 
\begin{eqnarray}
W_1 = A_{1}{^\dagger} U A_1 \quad \text{and} \quad W_2 = A_{2}{^\dagger} U A_2, 
\end{eqnarray}

\noindent
which leads to 
\begin{eqnarray}
A_1 W_1 A_{1}{^\dagger} = A_2 W_2 A_{2}{^\dagger} = U 
\end{eqnarray}
\noindent
arriving at
\begin{eqnarray}
W_1=BW_2{B}^{\dagger}.
\label{eq:atp}
\end{eqnarray}
Since $A_1$ and $A_2$ are orthogonal transformation matrices, the product $B\,(=A_1^{\dagger}A_2)$ also appears as an orthogonal matrix ($B^{\dagger}B=A_2^{\dagger}A_1A_1^{\dagger}A_2=I$) and therefore, the Eq. \ref{eq:atp} ensures the path independence of the calculated observables originated from different diabatic PESs matrices ($W_1$ and $W_2$).~\citep{mukherjee_jpca117,joy_pccp,soumya_bz,Mukherjee_2024}

\subsection{Calculation of Photoemission Spectra}\label{sec:spec}
  
\noindent
Once the BBO based diabatic PESs and couplings of the titled system (TiO$_6^{8-}$) are in hand, we carry out multi-state multi-mode Time-Dependent Discrete Variable Representation (TDDVR) dynamics~\citep{SAJCP113,SAJCP121,SAIJQC105,SSJPCA118,souvik_irpc37,joy_jpca} (see Section S10 of the ESI) to calculate the photoemission (PE) spectra of TiO$_6^{9-}$ moiety. While solving the TDSE during dynamical calculations, the nuclear wavefunctions at different time are employed to compute the autocorrelation function ($C(t)$):
\begin{subequations}
	\begin{eqnarray}
	C(t) &=& \langle \Xi(t) | \Xi(0)\rangle \label{eq:autocor1}\\
	&=& \Big\langle \Xi^{\star}\Big(\frac{t}{2}\Big) 
	\Big| \Xi\Big(\frac{t}{2}\Big)\Big\rangle \label{eq:autocor2}.
	\label{Eq:Auto-Correlation}
	\end{eqnarray}
	\end{subequations} 
 
\noindent 
It is worth mentioning that the second expression (Eq. \ref{eq:autocor2}) is more accurate and computationally less expensive compared to the first one (Eq. \ref{eq:autocor1}), but it (Eq. \ref{eq:autocor2}) can only be used for real initial wavefunction and totally symmetric ($A_1$) Hamiltonian. The calculated autocorrelation functions are further used to generate the spectral intensity at peak positions with the help of the following Fourier Transformation:
	\begin{eqnarray}
	I(\omega) \propto \omega \int_{-\infty}^{\infty} C(t) \exp(i\omega t) dt, \label{eq:abs}
	\label{Eq:Fourier}
	\end{eqnarray}
where the integration time limit is taken as 0 to 100 fs for our present calculation. \\

\noindent 
Since the resolution of the spectrometer results into phenomenological broadening in the experimental 
spectral lines, the theoretical autocorrelation function needs to be multiplied with a damping 
time-dependent function:
	\begin{eqnarray}
	h(t)=\exp\Big[-\Big(\frac{|t|}{\tau}\Big)^{2}\Big] \label{eq:tdf}
	\label{Eq:damping}
	\end{eqnarray}
where the unit of $\boldsymbol{\tau}$ is considered as 10$^{-14}$ sec. Furthermore, while calculating the propagation of the wavefunction and the autocorrelation function within a finite time frame, artifacts known as the Gibbs phenomenon can arise in the spectrum. These artifacts can be reduced by multiplying the autocorrelation with the following function,
\begin{eqnarray}
	g(t)= \cos^{n}\Big(\frac{\pi t}{2T}\Big),
	\label{Eq:cos-function}
\end{eqnarray}
\noindent
where $n$ is an integer number set to $2$ in this context and \textit{T} denotes the length of propagation. Readers are requested to go through earlier articles~\citep{SSJPCA118,souvik_irpc37,soumya_tfb} to acquire in-depth knowledge about the calculation
of theoretical spectra by TDDVR formalism.

\subsection{Calculation of Ferroelectricity}\label{sec:Ferr}

\noindent
In case of TiO$_6^{8-}$ unit of BaTiO$_3$ crystal, we will utilize presently calculated \textit{ab initio} adiabatic PESs to analyze the structural distortion, employing an equation derived from the mean-field Hamiltonian, $H = H_{0} - \lambda Q \langle Q \rangle$. This equation incorporates the potential energy, $U = U_{0} - \lambda Q \langle Q \rangle$, where $\langle Q \rangle$ is a parameter leading to a system of coupled transcendental equations~\cite{bersuker_cm5}:
\begin{eqnarray}
	\langle Q_{j}\rangle &=& \frac{\int Q_{j} \exp^{-U(Q)/{kT}} d^{3}Q}{Z}, \qquad Z= \int \exp^{-U(Q)/{kT}} d^{3}Q,
	\label{Eq:Partition_function}
\end{eqnarray}

\noindent
where $j$ represents one of the any axis ($x$ or $y$ or $z$) of distortion. Since this phase transition occurs due to the change of geometry from tetragonal to cubic or the reverse, one can consider that the titanium ion ($Ti^{4+}$) undergoes displacement along a specific direction, let say, the z-axis. Therefore, assuming zero displacements along the x and y axes ($Q_{x}=Q_{y}= 0$), the partition function for the displacement along the z-axis can be expressed as:

\begin{eqnarray}
	\langle Q_{z}\rangle &=& \frac{\int Q_{z} \exp^{-U(Q_{z})/{kT}} dQ_{z}}{Z}, \qquad Z= \int \exp^{-U(Q_{z})/{kT}} dQ_{z}.
	\label{Eq:Partition_function1}
\end{eqnarray}

\noindent
This framework is used to calculate the ferroelectric properties based on the structural distortions.

\section{Details of \textit {Ab Initio} calculation on TiO$_6^{8-}$ moiety}\label{sec:abinitio}

\noindent
In the present work, first we compute the vibrational frequencies of TiO$_6^{8-}$ unit of BaTiO$_3$ cluster with the aid of Coupled-Cluster with Singles and Doubles (CCSD) method employing atomic natural orbital (ANO-R1) basis set as implemented in MOLPRO quantum chemistry software~\citep{MOLPRO_brief} and compare those with the existing theoretical data~\cite{Bersuker_2015,A_Mahmoud} (see Table \ref{tab:thresh}). On the other hand, within the Franck-Condon (FC) domain of nuclear CS, the adiabatic PESs are calculated using Complete Active Space Self-Consistent Field (CASSCF) followed by Multi-Reference Configuration Interaction (MRCI) approach, where NACTs are evaluated by  CASSCF based Numerical Finite Difference (DDR) methodology. Since the degeneracies of the valence molecular orbitals ($t_{1u}$) as well as all other inner-shell orbitals are perfectly reproduced while taking all the six $Ti-O$ bonds equal with length of 2.0 \r{A}, such undistorted geometry is considered as the origin of the nuclear CS. Moreover, such choice of bond length (2.0 \r{A}) is in conformity both with experimental finding~\cite{Kwei_Bond_order_exp} and other theoretical prediction~\cite{Heliyon_bond_order_theor}. It is worth mentioning that the entire calculation is performed using  six (6) electrons in six (6) orbitals forming complete active space (CAS) involving filled HOMO ($t_{1u}$) and virtual LUMO ($t_{2g}^*$) orbitals leading to the correct order of energies for the ground ($^1A_{1g}$) and excited ($^1T_{1u}$) electronic states as well as degeneracies in the excited ($^1T_{1u}$) electronic states for the undistorted geometry ($Ti-O$ bond length = 2.0 \r{A}). \\

\noindent
Since the ground and first three excited states belong to $A_{1g}$ and $T_{1u}$ symmetry, respectively, the symmetry driven condition depicts that the $g$ symmetric triply degenerate normal modes ($t_{2g}$, frequency = 266.81 cm$^{-1}$) lead to JT distortion for the $T_{1u}$ symmetric excited electronic states, whereas PJT couplings are observed between $A_{1g}$ and $T_{1u}$ states along $t_{1u}$ normal modes (frequency = 123.87 cm$^{-1}$). The detailed conditions for the JT and PJT interactions between various electronic states and normal modes are analysed in Section S1 of the ESI. In our calculation, all the six normal modes (three $t_{1u}$ and three $t_{2g}$ symmetric) are included to construct six two-dimensional (2D) nuclear planes, namely $Q_{t_{1ux}}-Q_{t_{1uy}}$, $Q_{t_{1uy}}-Q_{t_{1uz}}$, $Q_{t_{1uz}}-Q_{t_{1ux}}$, $Q_{t_{2gx}}-Q_{t_{2gy}}$, $Q_{t_{2gy}}-Q_{t_{2gz}}$ and $Q_{t_{2gz}}-Q_{t_{2gx}}$ to compute the adiabatic PESs and NACTs followed by construction of diabatic Hamiltonian (PESs and couplings). We explore the contribution of both types of non-adiabatic couplings, namely, JT interactions in excited states and PJT stabilization of ground state, and finally, in the diabatic Hamiltonian through the photoemission (PE) spectra. Moreover, we also execute {\it ab initio} calculations of the 2D cross-section for the nuclear plane, $Q_{t_{1ux}}-Q_{t_{1uy}}$ for various values of $Q_{t_{1uz}}$ normal mode and obtain a global minima, where the corresponding bond lengths are presented in Table \ref{tab:thresh} along with other results obtained from Green's function DFT calculations~\cite{Bersuker_2015}. It may be noted that Green's function DFT~\cite{Bersuker_2015} calculations is based on single-reference, whereas we perfomed multi-reference (MRCI-SD) calculattion and thereby, the accuracy of our calculation is much higher.\\

\begin{table}[!htp]
	\caption{Comparison of various molecular parameters between presently calculated  and other theoretical results, where $Q_0$ represents the saddle point, $K_0$ is the force constant for the $Q$ displacement and global minimum  point, \{$Q_{t_{1ux}}^{gm}$, $Q_{t_{1uy}}^{gm}$, $Q_{t_{1uz}}^{gm}$\}.}
	\label{tab:thresh}
	\begin{center}
		\begin{small}
			\begin{tabular}{|c|c|c|}
		    	\hline
				 Parameters& CCSD & Green's Function DFT~\cite{Bersuker_2015}  \\
				\hline
				$\hbar w_{e}$ ($t_{1u}$) ($cm^{-1}$) & 123.87  & 193 (also Ref.~\citen{A_Mahmoud}) \\
				$\hbar w_{e}$ ($t_{2g}$) ($cm^{-1}$) & 266.81 &---  \multirow{3}{*}{}  \\
				\cline{1-2}
				& MRCI(SD) & \multirow{4}{*}{} \\
				\cline{1-2}
				$Q_{t_{1ux}}^{gm}$= $Q_{t_{1uy}}^{gm}$= $Q_{t_{1uz}}^{gm}$ ($\AA$) & 0.51 & 0.14  \\
				
				$Q_o$ ($\AA$) & 0.64 &0.16 \\
				
				$K_o$ (eV/\AA) & 44.18 &55 \\

				$E_{PJT}$ ($eV$) &  2.582& 4.0 \\
				$E_{JT}$ ($eV$) &  0.0182& --- \\
				\hline
				
			\end{tabular}
		\end{small}
	\end{center}
\end{table}

\section{ \textit{Ab Initio} and ADT Quantities}\label{sec:adt}

\begin{figure}[!htp]
	\centering
	\includegraphics[width=0.9\linewidth]{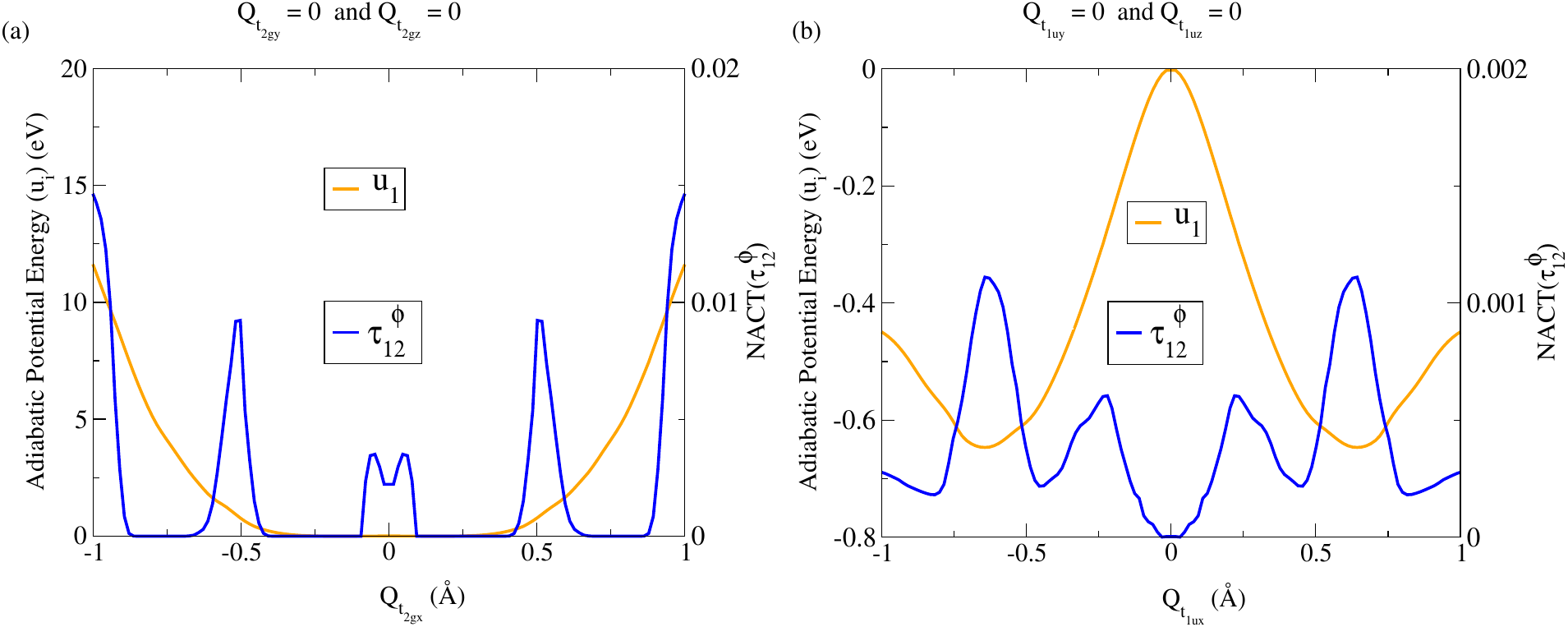}
     \caption{For TiO$_6^{8-}$ unit, 1D curves of lowest adiabatic PES ($u_1$) and the associated NACT ($\tau_{12}^{\phi}$) are presented along (a) $Q_{t_{2gx}}$ and (b) $Q_{t_{1ux}}$ normal modes keeping the other normal modes fixed at zero (0).}
	\label{fig:t1g_1D}
\end{figure}

\noindent
Since we intend to acquire the structural insights of TiO$_6^{8-}$ unit of BaTiO$_3$ crystal, the 
inherent non-adiabatic interactions (JT/accidental CIs and PJT couplings) over six chosen normal mode pairs are explored. In order to investigate the validity of quantization of NACTs along a closed path over the 2D nuclear planes, mass-weighted normal mode coordinates are converted into their polar analogues as depicted below:
\begin{eqnarray}
Q_{k} = \rho \cos \phi \quad \text{and} \quad Q_{l} = \rho \sin \phi, \nonumber
\end{eqnarray}
\noindent
where $Q_k$, $Q_l$ are any two components of $t_{1u}$ or $t_{2g}$ symmetric normal modes, and $\rho$ and $\phi$ symbolize their polar counterpart. For better identification, the lowest four electronic states are numbered as per the increasing order of energy: $^1A_{1g} \rightarrow 1\,\text{and}\, ^1T_{1u} \rightarrow 2-4$. For each 2D nuclear plane ($\rho-\phi$), we have considered 70 $\times$ 181 \textit{ab initio} grid points, where the $\rho$ and $\phi$ coordinates span from 0.2 to 14.0 and 0 to $2\pi$, respectively. In other words, the normal modes ($\lbrace Q_i\rbrace$) range from -1.5 \AA\, to 1.5 \AA, which encapsulate the entire FC domain of nuclear CS. Since the displacements along all the three components ($x$, $y$ or $z$) of $t_{1u}$ or $t_{2g}$ symmetric normal modes lead to identical geometries, the adiabatic potential energy curves (PECs) and NACTs are superimposable with each other (see in Figures \ref{fig:t1g_1D} and S1 of the ESI). One can find repetition of those profiles every after 90$^o$ in the polar forms of the normal modes, $Q_x$= $\rho \ cos\phi$ and $Q_y$= $\rho \ sin\phi$ ($ 0\le \phi < 2\pi $), which is the inherent  symmetry of the molecular species. Therefore, the detailed discussion on adiabatic PESs, NACTs, ADT angles, ADT matrices and diabatic PESs matrices are presented below only for two representative planes, $Q_{t_{2gx}}-Q_{t_{2gy}}$ and $Q_{t_{1ux}}-Q_{t_{1uy}}$ belonging to two different symmetries. 

\newpage

\subsection{$Q_{t_{2gx}}-Q_{t_{2gy}}$ Pair}\label{ssec:pair1}

\begin{figure}[!htp]
	\centering
	\includegraphics[width=0.8\linewidth, height=0.4\textheight]{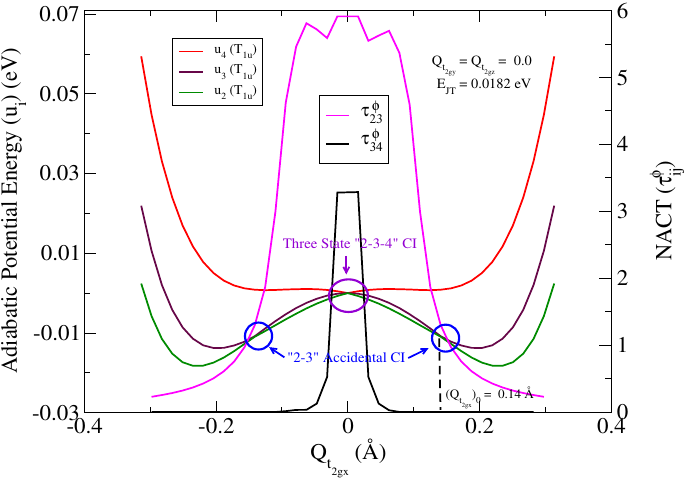}
	\caption{For TiO$_6^{8-}$ unit, 1D cuts of adiabatic PESs (u$_2$, u$_3$ and u$_4$) and NACTs ($\tau_{23}^{\phi}$ and $\tau_{34}^{\phi}$) for three excited electronic states ($T_{1u}$)  are shown along $Q_{t_{2gx}}$ normal modes keeping the other $t_{2g}$ coordinates fixed at zero (0). Similarly, the 1D cuts of adiabatic PESs and NACTs as functions of $Q_{t_{2gy}}$ or $Q_{t_{2gz}}$ will be exactly same.}
	\label{fig:t1g_en}
\end{figure}

\begin{figure}[!htp]
	\centering
	\includegraphics[width=1.0\textwidth]{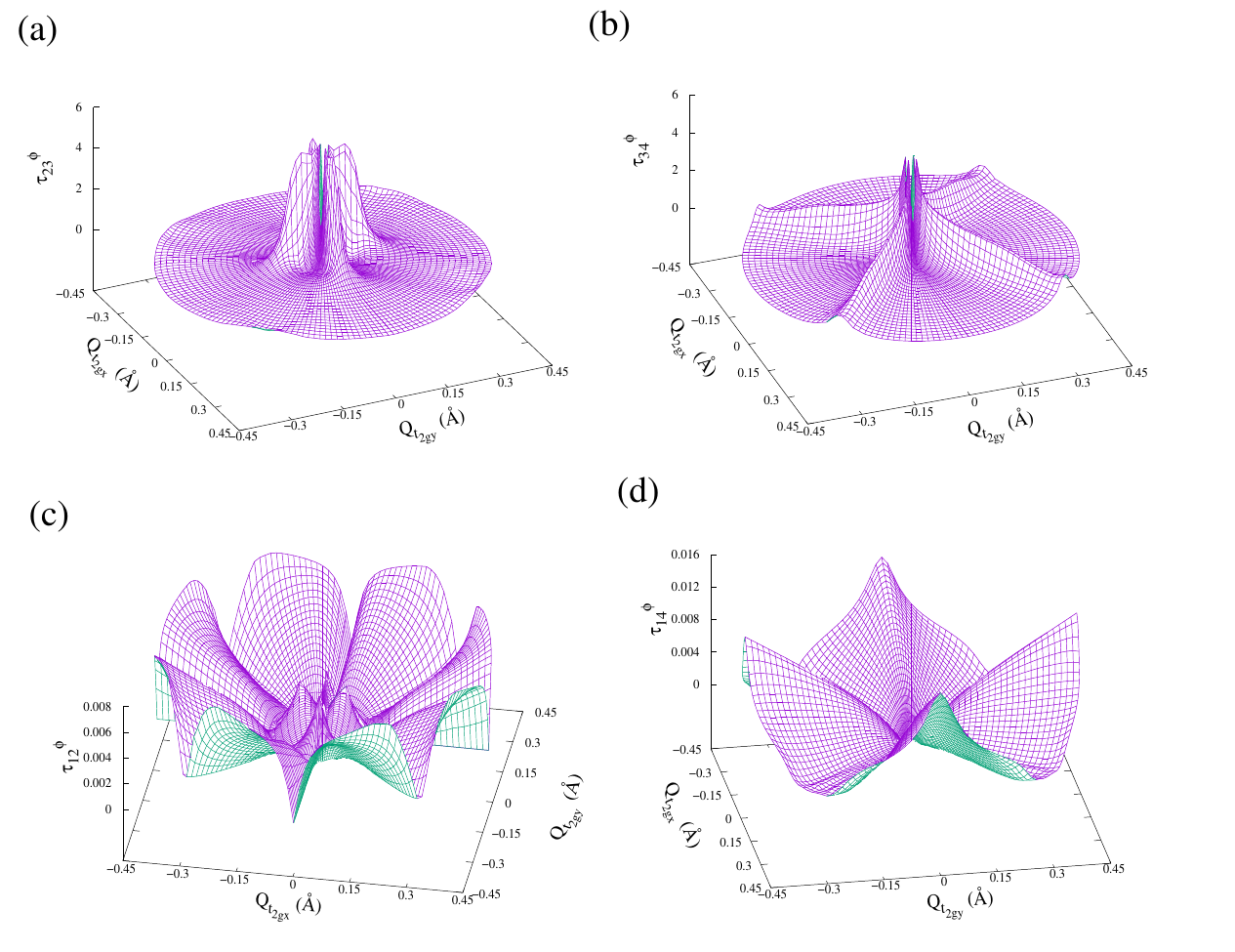}
	\caption{Variation of $\phi$ component of NACTs, namely, (a) $\tau^{\phi}_{23}$, (b) $\tau^{\phi}_{34}$, (c) $\tau^{\phi}_{12}$ and (d) $\tau^{\phi}_{14}$ are plotted over $Q_{t_{2gx}} - Q_{t_{2gy}}$ plane. The functional forms of NACT for four ``2-3"  accidental CIs and one three state (``2-3-4") JT CIs show coalesce of each type CIs at the centre of $t_{2g}$ coordinates.  }
	\label{fig:t1g_na}
\end{figure}

\noindent
The one-dimensional (1D) functional forms of adiabatic PESs (scaled with respect to the excited state energy at $\lbrace Q_0\rbrace$=0) and NACTs for three excited electronic states ($^1T_{1u}$)  are presented along $Q_{t_{2gx}}$ normal modes in Figure \ref{fig:t1g_en}, which clearly shows a three point (``2-3-4") JT CI at the origin ($\{$$Q_{t_{2gx}}$, $Q_{t_{2gy}}$ and $Q_{t_{2gz}}$$\}$ $=0$). On the other hand, two (2) additional ``2-3" accidental CIs are observed along each normal mode $Q_{t_{2gx}}$ = 0.14 \AA\ at $Q_{t_{2gy}}$ = 0 leading to total four (4) ``2-3" CIs over $Q_{t_{2gx}}-Q_{t_{2gy}}$ plane. Further details, including calculation of local topographic parameters and model double-cone adiabatic PESs around the ``2-3" CI, are provided in Section S5 of the ESI. These results confirm that the degeneracy is lifted linearly, consistent with the conical nature of the intersections. Even though $Q_{t_{2g}}$ is JT active mode for $^1T_{1u}$ symmetric electronic states, the magnitude of JT stabilization energy appears as very low (0.0182 eV). The 2D functional features of $\phi$ components of NACTs associated with JT and accidental CIs ($\tau^{\phi}_{23}$ and $\tau^{\phi}_{34}$) are presented over $Q_{t_{2gx}}-Q_{t_{2gy}}$ plane in Figures \ref{fig:t1g_na}(a-b). In case of ``2-3" interactions, singularities of NACTs for all the four accidental CI points coalesce to one connecting seam passing through the origin, whereas $\tau^{\phi}_{34}$ exhibits singularity only at the ``2-3-4" JT CI point. On the other hand, PJT couplings are observed between $A_{1g}$ and $T_{1u}$ electronic states (``1-2", ``1-3" and ``1-4") among which $\tau^{\phi}_{12}$ and $\tau^{\phi}_{14}$ are displayed in Figures \ref{fig:t1g_na}(c-d) to depict their well-behaved functional forms over the chosen domain of nuclear CS.\\

\newpage

\begin{figure}
	\centering
	\includegraphics[width=0.8\textwidth]{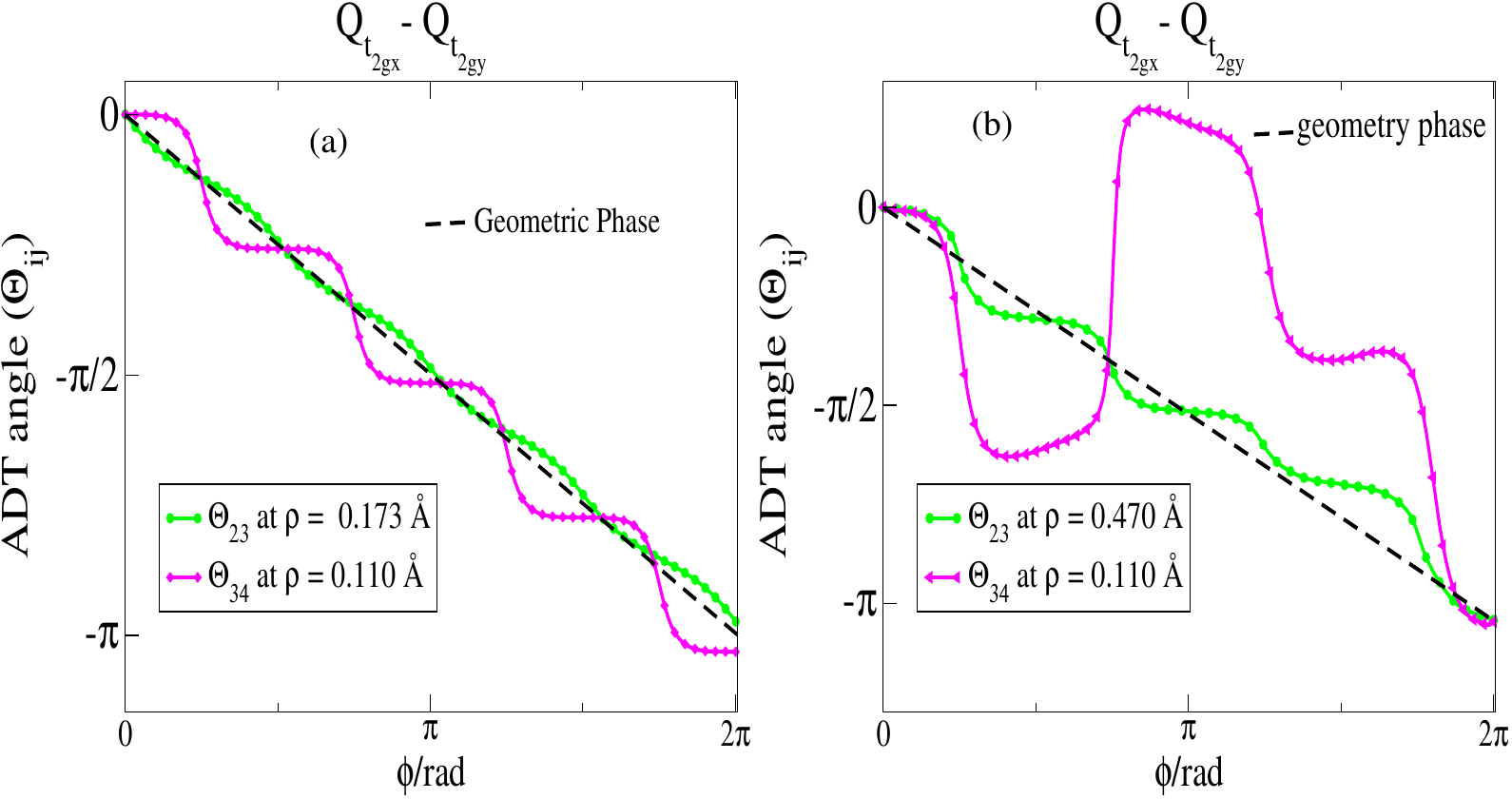}
	\caption{ 1D cuts of ADT angles for $Q_{t_{2gx}} - Q_{t_{2gy}}$ pairs, namely, $\Theta_{23}$ and $\Theta_{34}$ originated from (a) two and (b) four state ADT calculations are presented along the angular coordinate, $\phi$ for specific values of $\rho$, which attain the magnitude of $\pi$ at the end of the contour.}
	\label{fig:t1g_an}
\end{figure}

\noindent
As per Cauchy's residue theorem (see Eq. \ref{eq:quant}), if a closed contour of integration encloses PJT interaction(s) or JT or RT CI(s)/seam(s), the resulting ADT angles attain the magnitude of zero (0) or integer ($n$) multiple of $\pi$ or $2\pi$ and the associated diagonal elements of ADT matrices undergo $n$ number of sign inversions. The calculation of ADT angles employing the ADT equation, $\nabla A + \tau A = 0$, are carried out using the \textit{ab initio} calculated NACM ($\tau$) and thereby, there is no control on the nature of the profiles (see Eqs.~\ref{eq:adteq} and \ref{eq:ADTgen}). In fact, within numerical accuracy, one can say ADT angles are ``\textit{ab initio}" calculated and their functional nature could be linear or non-linear, but at the end of the closed contour, those angles attain 0, $n \pi$ and $n\cdot 2\pi$ depending upon the nature of the interaction, PJT, JT and RT, respectively. The geometric phase is linearly dependent on the circular coordinate ($\phi$), where the ADT angles could vary non-linearly as a function of such coordinates ($\phi$). On the other hand, while constructing the diabatic PESs and couplings, we solve the ADT equation along any chosen path and thereby, that is no unique diabatic matrix, but those matrices (due to different paths) are related through the orthogonal transformation providing unique observables~\citep{mukherjee_jpca117,joy_pccp,soumya_bz} (see Subsection \ref{sec:path}). On the contrary, while locating the CIs and their nature (PJT, JT and RT), we need to solve ADT equation along a circular coordinate ($\phi$) fixing the other coordinates at specific value and thereby, the functional form of ADT angle along such path will be specific for each different CIs. In the present work, diabatization is carried out using two state as well as four state ADT equations, where the quantization of NACTs is affirmed for both the cases. In case of $Q_{t_{2gx}} - Q_{t_{2gy}}$ pair, Figure \ref{fig:t1g_an} depicts 1D variation of ADT angles ($\Theta_{23}$ and $\Theta_{34}$) along $\phi$ coordinate for specific values of $\rho$, which attain the magnitude of $\pi$ for two and four states. It is worth mentioning that the ADT angles appear as non-linear complicated functions of $\phi$, which differ from two to four state calculation even for a specific molecular process indicating four state sub-Hilbert space. The non-linear functional form of each ADT angle leads to the associated phase ($\omega_{\phi}(\phi)$) topological ($e^{\pm i\omega_{\phi}(\phi)}$) in nature instead of geometric ($e^{\pm i\frac{n\phi}{2}}$, $n$ is the number of enclosed CI(s)/seam(s))~\citep{herzberg_dfs35,pancharatnam_pias44,berry_prca392}.

\vspace{0.2 cm}

\vspace{0.2 cm}

\begin{figure}[htp]
\centering
\includegraphics[width=0.9\textwidth]{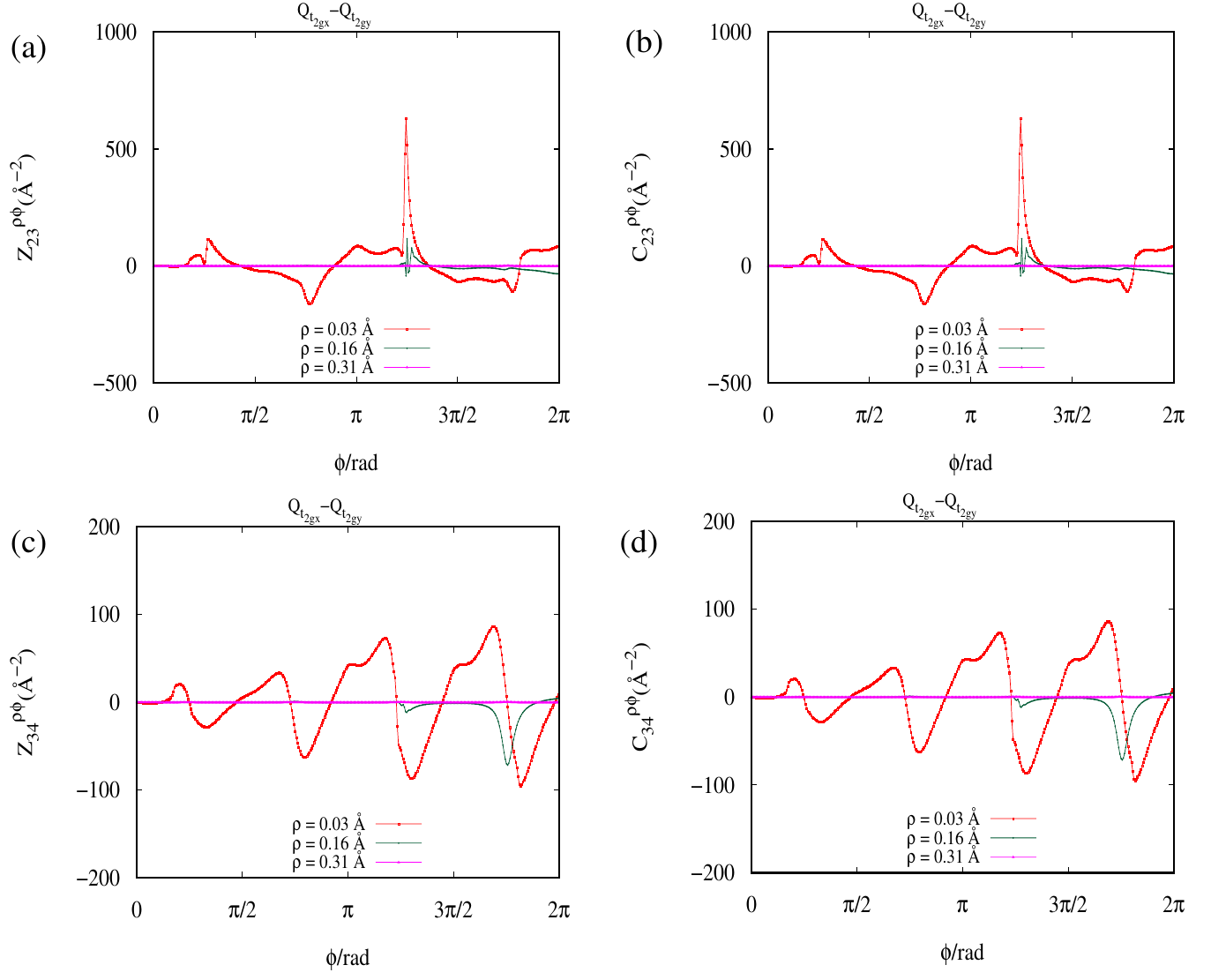}
\caption{The left and right panels depict Mathematical Curls [(a) $Z^{\rho\phi}_{23}$ and (c) $Z^{\rho\phi}_{34}$] and ADT Curls [(b) $C^{\rho\phi}_{23}$ and (d) $C^{\rho\phi}_{34}$] along the angular coordinate ($\phi$) at three different $\rho$ values for $Q_{t_{2gx}}-Q_{t_{2gy}}$ plane of TiO$_6^{8-}$ unit of BaTiO$_3$ crystal. }
\label{fig:t1g_curl}
\end{figure}

\noindent
Once the ADT angles are in hand, we employ those four state ADT angles in the explicit expressions of NACTs (see Section S3.2 of the ESI) to recalculate those quantities to evaluate Mathematical Curls ($Z^{\rho\phi}_{ij}$) and ADT Curls ($C^{\rho\phi}_{ij}$).~\citep{mantu_curl} Figure \ref{fig:t1g_curl} shows that the $Z^{\rho\phi}_{23}$ and $C^{\rho\phi}_{23}$ (which incorporates ``2-3" accidental JT CIs) as well as $Z^{\rho\phi}_{34}$ and $C^{\rho\phi}_{34}$ (which accommodates symmetry driven ``3-4" JT CI) are almost superimposed with each other and therefore, the resulting matrix elements of Curl Condition ($F^{\rho\phi}_{23}$ and $F^{\rho\phi}_{34}$) appear close to zero ($< 10^{-15}$ $\AA^{-2}$). In other words,  those four electronic states ($^1A_{1g}$, $^1T_{1u}$) of TiO$_6^{8-}$ unit form a subspace. 

\vspace{0.2 cm}

\begin{figure}[htp]
\centering
\includegraphics[width=0.8\textwidth]{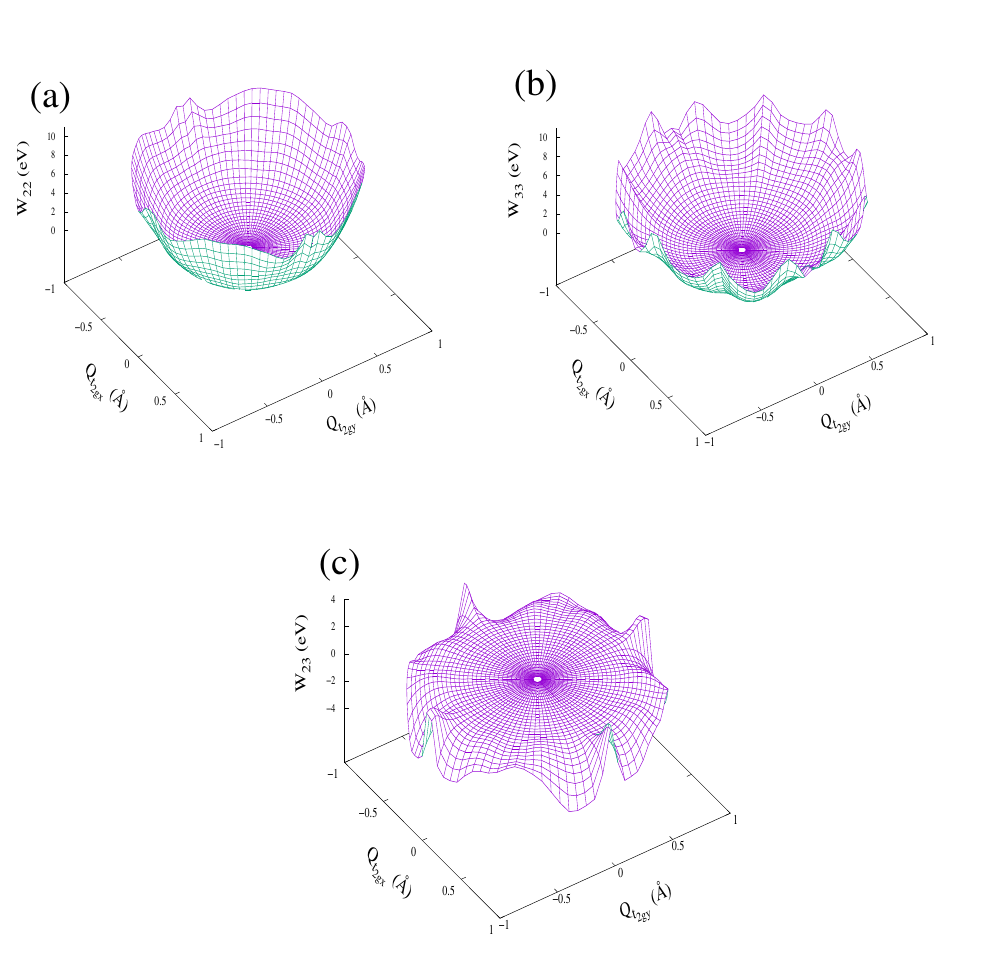}
\caption{2D functional forms of diabatic surfaces for $Q_{t_{2gx}}-Q_{t_{2gy}}$ plane of TiO$_6^{8-}$ unit of BaTiO$_3$ crystal, namely, (a) $W_{22}$ and (b) $W_{33}$, and the associated diabatic coupling, (c) $W_{23}$ are presented, where all the quantities appear as well-behaved functions of nuclear coordinates. }
\label{fig:t1g_dia}
\end{figure}

\noindent
While computing the diabatic Hamiltonian over 2D nuclear planes, the ADT equations are first integrated along $\phi$ coordinate from $0\; \text{to}\; 2\pi$ at a fixed magnitude of $\rho$ $(=\rho_{min})$ and then, along $\rho$ coordinate for each $\phi$ value (Path II, see Section S9 of the ESI). Once we obtain the ADT angles vis-\'a-vis ADT matrices, the similarity transformation (see Eq. \ref{eq:diaben}) is carried out to obtain the diabatic PESs matrices. Figure \ref{fig:t1g_dia} displays some representative diabatic PESs (W$_{22}$ and W$_{33}$) and coupling (W$_{23}$) over the $Q_{t_{2gx}} - Q_{t_{2gy}}$ plane. It is interesting to note that the diabatic PESs and couplings exhibit a four-fold symmetry over the nuclear plane, since nuclear displacements along each of the normal modes, $Q_{t_{2gx}}$ and $Q_{t_{2gy}}$ lead to equivalent geometry as depicted in Figures \ref{fig:t1g_1D} and S1 of the ESI. Such diagrams clearly depict smooth, continuous and single-valued forms of diabatic PES matrix elements at each and every grid point over the nuclear CSs, which validates the workability of our first principle based BBO formalism for a transition metal complex (solid state perovskite) possesing
four electronic state sub-Hilbert space.

\subsection{$Q_{t_{1ux}}-Q_{t_{1uy}}$ Pair}\label{ssec:pair2}

\begin{figure}[htp]
\centering
\includegraphics[width=0.8\textwidth]{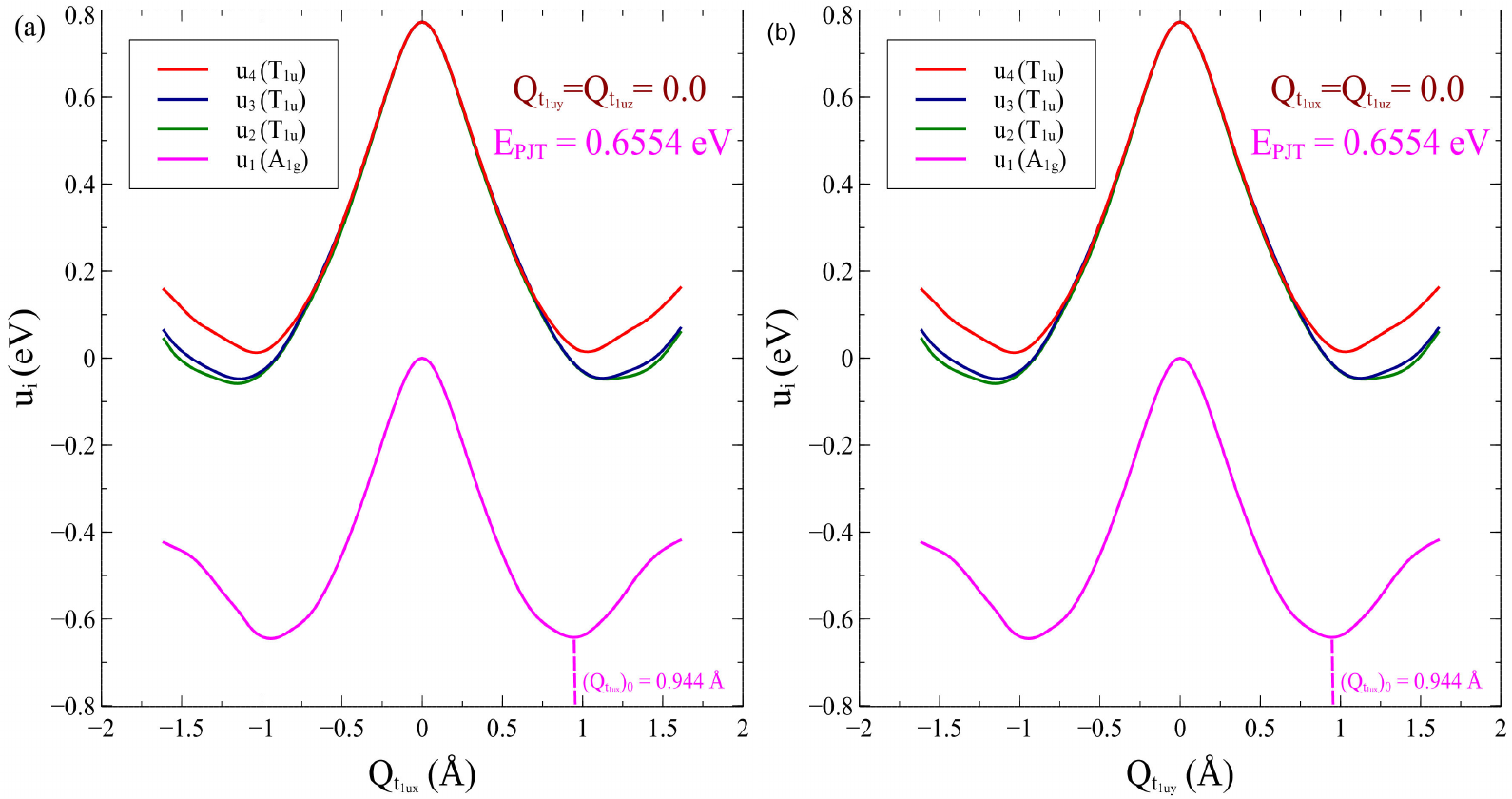}
\caption{1D curves of adiabatic PESs for ground ($A_{1g}$) and first three excited electronic states ($T_{1u}$) are presented along (a) $Q_{t_{1ux}}$ and (b) $Q_{t_{1uy}}$ normal mode keeping the other $t_{1u}$ coordinates ((a) $Q_{t_{1uy}}$ or (b) $Q_{t_{1ux}}$ and $Q_{t_{1uz}}$) fixed at zero (0).}
\label{fig:t1u_adia_1D}
\end{figure}

\noindent
The $u$-symmetric normal mode pairs, namely, $Q_{t_{1ux}}-Q_{t_{1uy}}$ are mainly responsible for profound PJT interactions between ground
($A_{1g}$) and first excited ($T_{1u}$) electronic states. The 1D curves of the four adiabatic PESs ($u_1$, $u_2$, $u_3$ and $u_4$) are presented in Figure \ref{fig:t1u_adia_1D} along  $Q_{t_{1ux}}$ and $Q_{t_{1uy}}$ normal modes indicating the presence of minima around $Q_{t_{1ux}}$ = $\sim$ 0.944 \AA\, or $Q_{t_{1uy}}$ = $\sim$ 0.944 \AA. In case of first three excited states, the degeneracies remain almost intact throughout the 1D curve due to absence of any JT distortion.

\begin{figure}[htp]
	\centering
	\includegraphics[width=1.0\linewidth, height=0.5\textheight]{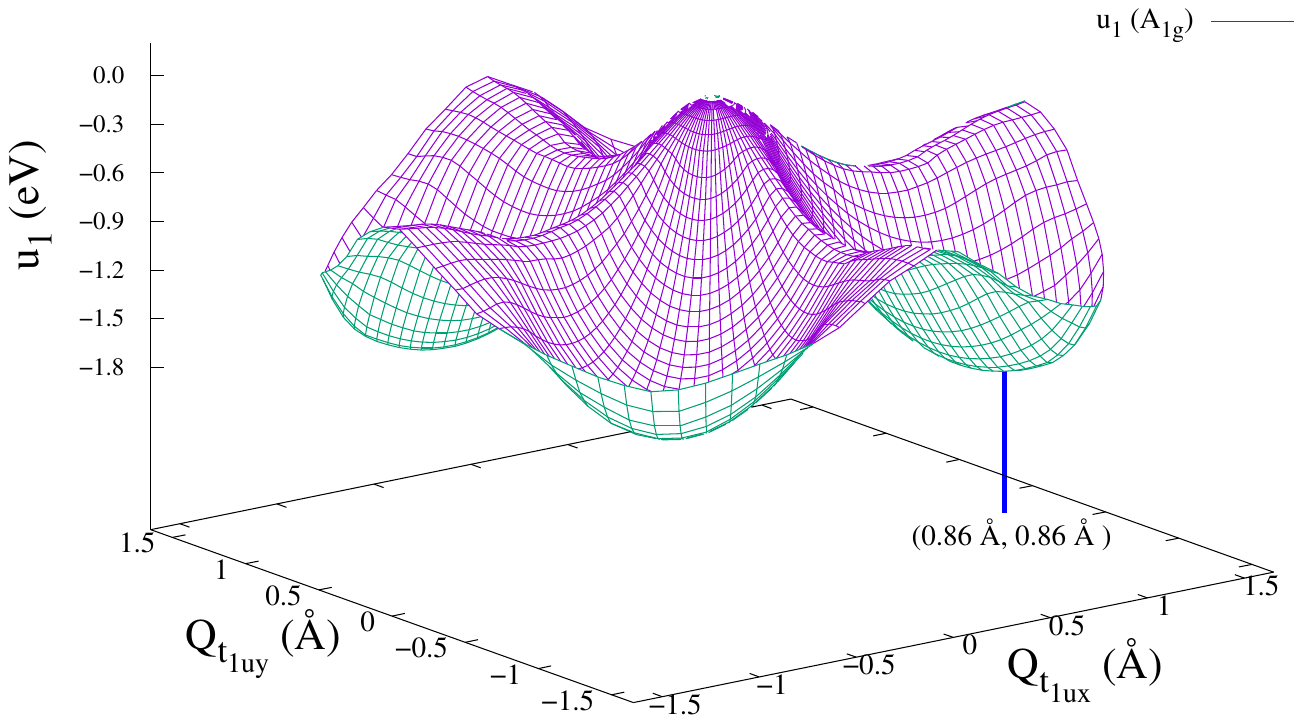}
	\caption{2D functional form of the ground adiabatic surface, $u_{1}$ is presented over the $Q_{t_{1ux}}-Q_{t_{1uy}}$ plane, where the nuclear configuration of PJT distorted minima is shown at $Q_{t_{1ux}}$ = $Q_{t_{1uy}}$ = 0.86 \AA.}
	\label{fig:t1u_adia}
\end{figure}

\noindent
On the other hand, Figure \ref{fig:t1u_adia} depicts the 2D PESs over $Q_{t_{1ux}}-Q_{t_{1uy}}$ plane,
which leads to four equivalent elongated octahedron structure (all the four positive and negative combinations, $Q_{t_{1ux}}$ = $Q_{t_{1uy}}$ = $\pm$  0.86 \AA ) with high stabilization energy ($\sim$ 1.578 eV). On the other hand, if all the three normal modes vary simultaneously, one can find eight global minima appearing at $Q_{t_{1ux}}$ = $Q_{t_{1uy}}$ = $Q_{t_{1uz}}$ = $\pm $0.51 \AA\ stabilized by 2.582 eV, which differs substantially from predicted value ($\sim$4 eV) by Green's function DFT methdology~\cite{Bersuker_2015}. The accuracy of present calculation for PJT stabilization energy is validated through the following investigations.

\vspace{0.2 cm}

\begin{figure}[!htp]
	\centering
	\includegraphics[width=1.0\linewidth, height=0.6\textheight]{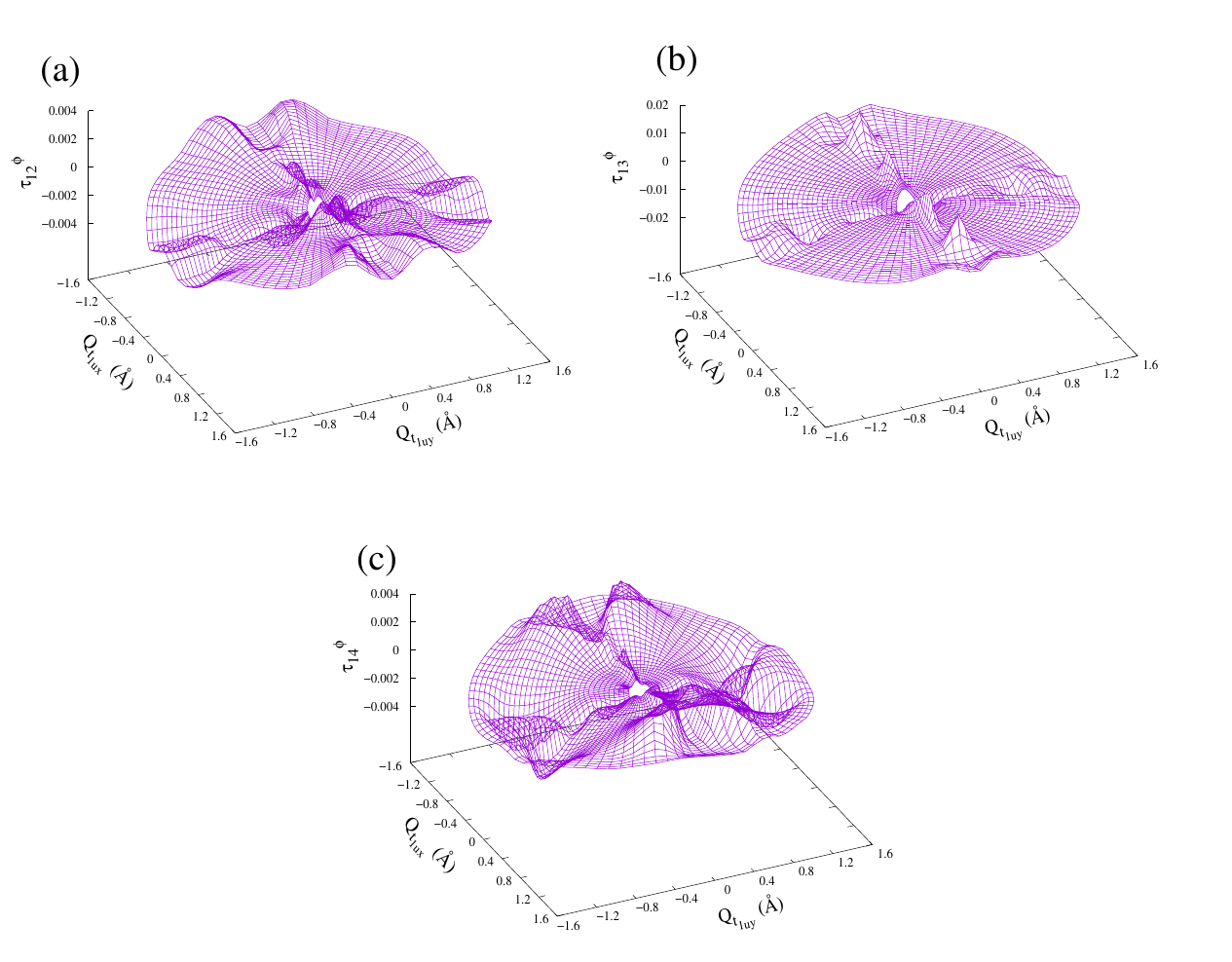}
	\caption{2D variation of $\phi$ component of NACTs, namely, (a) $\tau^{\phi}_{12}$, (b) $\tau^{\phi}_{13}$ and (c) $\tau^{\phi}_{14}$ are presented over $Q_{t_{1ux}} - Q_{t_{1uy}}$ plane. For all cases, the NACT profiles do not exhibit any singularity over the chosen domain of nuclear CS.}
	\label{fig:t1u_na}
\end{figure}

\noindent
In Figure S4 of the ESI, we depict the triply degenerate ($t_{1u}$) bending modes, where Figure S5 of the ESI displays the initial as well as optimized geometries. The Figure S6 of the ESI shows the PJT stabilized ground state ($A_{1g}$) in which the $Ti-O$ bond lengths along the \{\emph{x-y}\} and \emph{z} directions are 2.07 \r{A} and 2.0 \r{A}, respectively and the bond angles are close to 104$^{o}$ and 75$^{o}$. To verify such structure, we have carried out geometry optimization calculations using relaxed scans at the SA-4-CAS (6o,6e)/ANO-R1 level of theory, followed by a single-point MRCI energy calculation. On the process of performing such  optimization, the initial guess geometry has been taken as \{$Q$ = 0\}, which represents a perfect octahedron configuration with $Ti-O$ bond length,  2.0 \r{A}. In Figure S5 of the ESI, the optimized geometry displays elongated $Ti-O$ bond length, 2.07 \r{A}, along the  \{\emph{x-y}\}  direction, but the length along the \emph{z}-direction remains same, 2.0 \r{A}. Moreover, the bond angles are very close to 90$^{o}$. It is worth noting that any other initial guess geometries (including elongated octahedron structure) also lead to the same optimized structure. In other words, the ground state minimum energy structure obtained by scanning along the $t_{1u}$ mode closely matches the optimized structure from the relaxed scan. Furthermore, Figure \ref{fig:t1u_adia} shows that the PJT stabilization energy of the optimized structure is 1.5 eV, which coincides with the stabilization energy obtained from a 2D scan over the $t_{1u}$ normal mode pair. Since the $u$-symmetric normal modes result into only PJT couplings (no JT interaction) within the low-lying four electronic states of TiO$_6^{8-}$ unit, the 2D functional forms of $\phi$ components of NACTs representing (a) ``1-2" ($\tau^{\phi}_{12}$), (b) ``1-3" ($\tau^{\phi}_{13}$) and (c) ``1-4" ($\tau^{\phi}_{14}$) PJT interactions are depicted in Figure \ref{fig:t1u_na}, which naturally devoid of any singularity.

\vspace{0.2 cm}

\begin{figure}[!htp]
	\centering
	\includegraphics[width=0.5\textwidth]{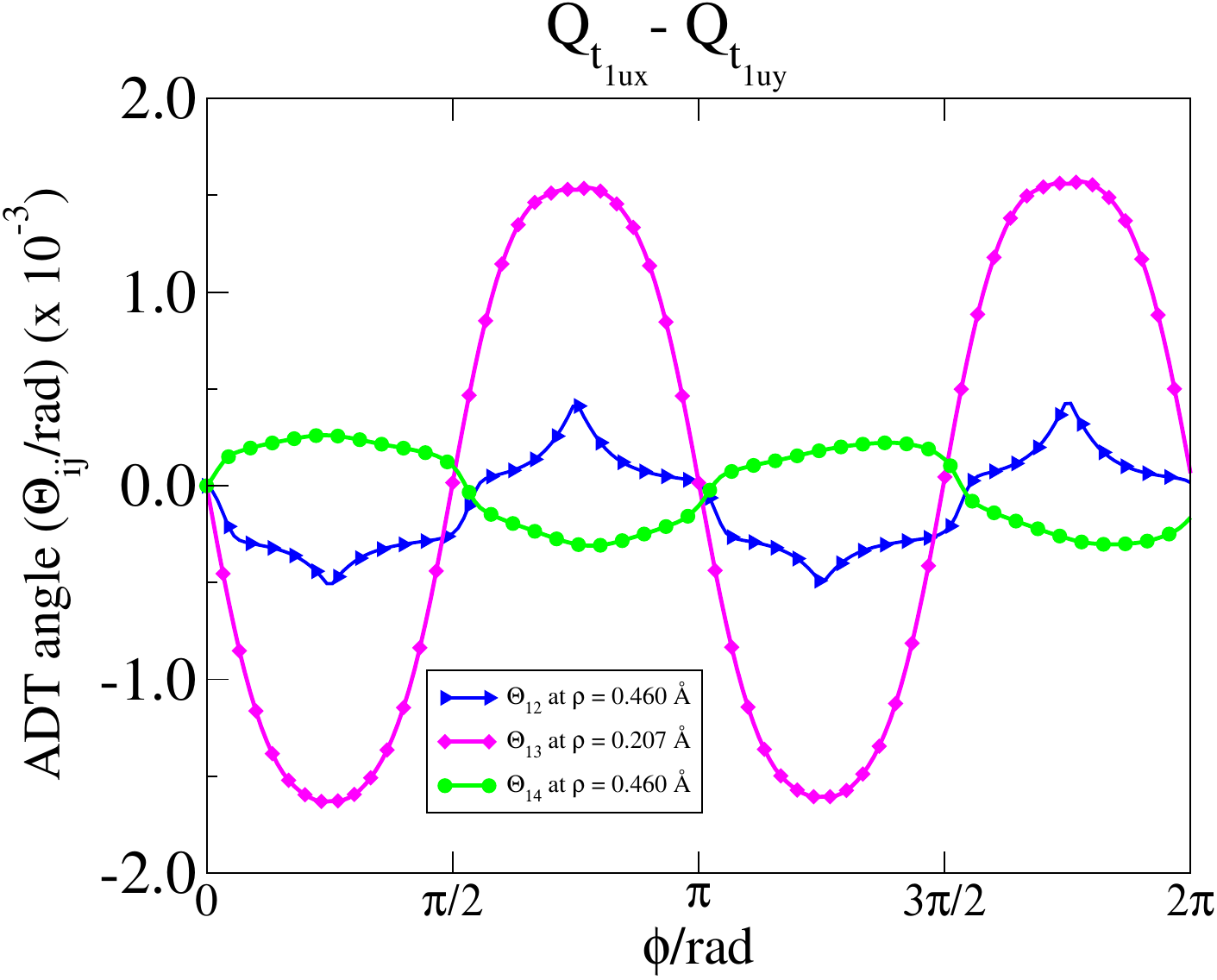}
	\caption{ 1D variation of ADT angles, namely, $\Theta_{12}$, $\Theta_{13}$ and $\Theta_{14}$ for $Q_{t_{1ux}} - Q_{t_{1uy}}$ pairs are presented along the angular coordinate, $\phi$ for fixed values of $\rho$ acquiring the magnitude of zero (0) at the end of the contour.}
	\label{fig:t1u_an}
\end{figure}

\noindent
As depicted by Cauchy's residue theorem (see Eq. \ref{eq:quant}), if a closed contour encapsulates PJT interaction(s), the calculated ADT angles acquire the value of zero (0) and therefore, the corresponding diagonal elements of ADT matrices do not undergo any sign inversion. Figure \ref{fig:t1u_an} depicts 1D functional forms of three ADT angles, namely, $\Theta_{12}$, $\Theta_{13}$ and $\Theta_{14}$ along the $\phi$ coordinates at specific fixed values of $\rho$. It is interesting to observe that the profiles of ADT angles attain zero (0) value at the end of the contour, but acquire non-zero magnitude in the intermediate $\phi$ values.\\

\newpage

\begin{figure}[htp]
	\centering
	\includegraphics[width=0.8\linewidth, height=0.5\textheight]{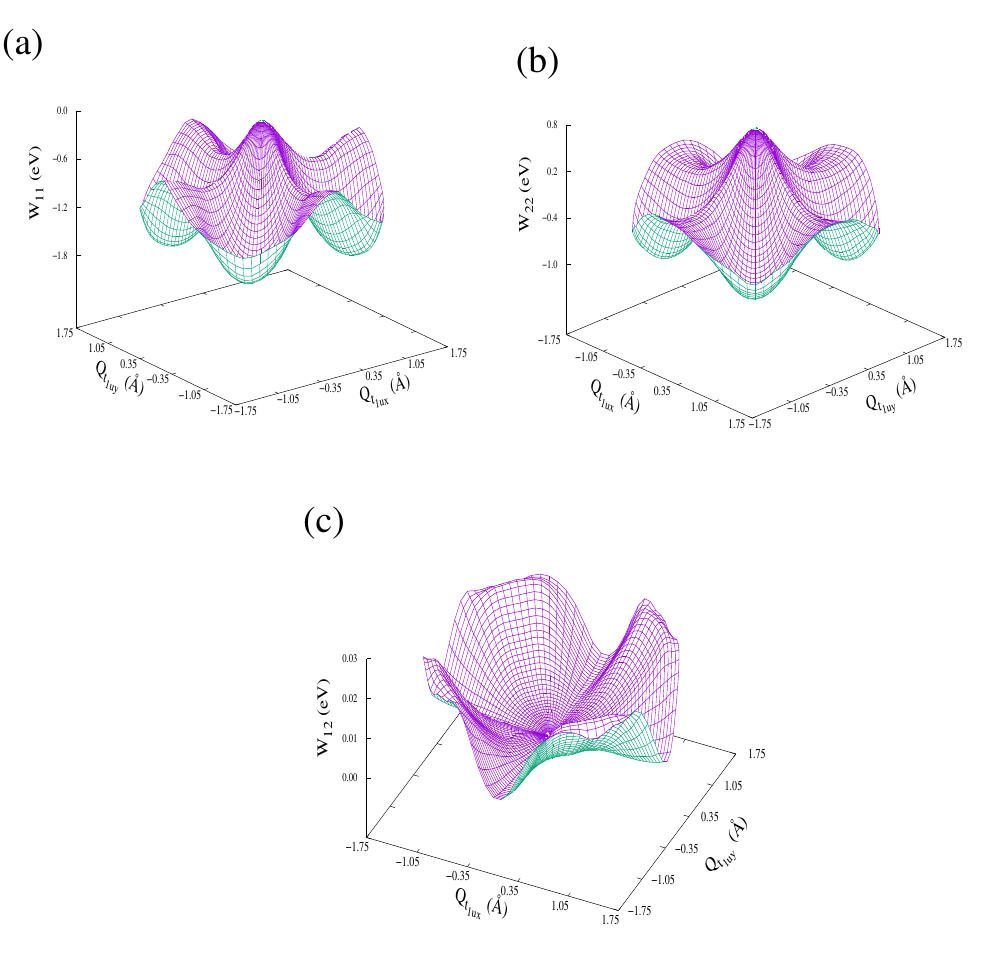}
	\caption{For $Q_{t_{1ux}}-Q_{t_{1uy}}$ plane of TiO$_6^{8-}$ unit of BaTiO$_3$ crystal, 2D functional forms of diabatic surfaces, namely, (a) $W_{11}$ and (b) $W_{22}$, and the associated diabatic coupling, (c) $W_{12}$ are presented, where all the quantities appear as well-behaved functions of nuclear coordinates. }
	\label{fig:t1u_dia}
\end{figure}

\noindent
For $Q_{t_{1ux}}-Q_{t_{1uy}}$ pair of normal modes, path II (see Section S9 of the ESI) is employed to compute the diabatic Hamiltonian
over 2D nuclear planes. While carrying out such calculation, the computed ADT matrices are employed in the similarity transformation (see Eq. \ref{eq:diaben}) to obtain the diabatic PESs matrices. Some representative diabatic PESs (W$_{11}$ and W$_{22}$) and coupling (W$_{12}$) are depicted in Figure \ref{fig:t1u_dia} over the  nuclear plane, $Q_{t_{1ux}} - Q_{t_{1uy}}$. 

\newpage

\section {Observables: Theoretical and Experimental Results}

\noindent
 The literature on experimental evidence of electronic transition leading to satellite peak in dielectric function is absent, which may be due to the negligibly small JT stabilisation energy (0.0182 eV) arising from the JT-splitted triply degenerate electronic state $T_{1u}$. The experimental prediction of ferroelectricity of BaTiO$_3$ has been explored using the pseudo Jahn-Teller effect on ground electronic states through $A_{1g} - T_{1u}$ interaction, that has been studied in present and other theoretical calculation also. We employed \textit{ab initio} calculated adiabatic PES and BBO based diabatic Hamiltonian to depict the nature of spontaneous polarisation and photoemission spectra, respectively.

\subsection{Photoemission Spectra} \label{Photoemission Spectra}

\begin{figure}[!htp]
	\centering
	\includegraphics[width=0.3\linewidth, height=0.5\textheight]{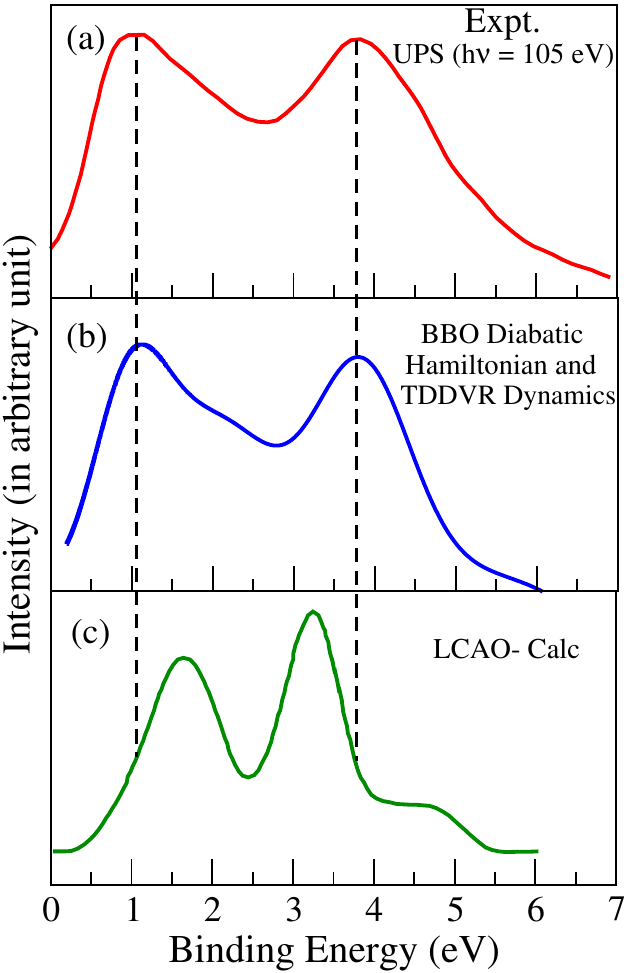}
	\caption{Comparison between experimental valence electronic spectra and theoretical spectra: (a) UPS spectra of BaTiO$_3$ single crystal measured by Cord and coworkers~\cite{R_Cord_1985}, (b) present work and (c) calculation of bulk BaTiO$_3$ by Pertosa {\it et al.}~\cite{Pertosa_Theory_Spectra}} 
	\label{fig:sp}
\end{figure}

We intend to probe the Jahn-Teller (JT) and pseudo Jahn-Teller (PJT) effect arising from electron-nuclear couplings in photoemission (PE) spectra, specifically engaging the diabatic PESs of transition metal oxide TiO$_6^{8-}$  constructed with octahedral as well as all possible distorted geometries. In this investigation, we explore how the ground (\textit{A$_{1g}$}) and excited (\textit{T$_{1u}$}) electronic states of TiO$_6^{8-}$ are influenced due to the interaction through the nuclear motion of the $t_{1u}$ and $t_{2g}$ normal modes or vice versa, where six two-dimensional (2D) nuclear planes, namely $Q_{t_{1ux}}-Q_{t_{1uy}}$, $Q_{t_{1uy}}-Q_{t_{1uz}}$, $Q_{t_{1uz}}-Q_{t_{1ux}}$, $Q_{t_{2gx}}-Q_{t_{2gy}}$, $Q_{t_{2gy}}-Q_{t_{2gz}}$ and $Q_{t_{2gz}}-Q_{t_{2gx}}$ are considered in the diabatic Hamiltonian. Since the electron-nuclear couplings are explicitly incorporated into the diabatic PESs matrices of TiO$_6^{8-}$, those PESs matrices are used to carry out the TDDVR quantum dynamical calculations for elucidation of the PE spectra. The TDDVR basis set (grid points) for $t_{1u}$ and $t_{2g}$ normal modes are optimized using a wide range of basis functions starting from (3,3,3,3,3,3) [729 grid points] to (15,15,15,15,15,15) [1,13,90,625 grid points], where the set (13,7,7,13,11,11) [10,02,001 grid points] leads to converged PE spectral profile. In Section S11 of the ESI, a detailed discussion on the convergence analysis with suitable diagram is depicted in Figure S8. The quantum dynamical calculations are initiated by the wavepacket [$\Xi(0)$], modelled as a product of Gaussian wavepackets (GWPs) representing the normal modes ($t_{1u}$ and $t_{2g}$) of the TiO$_6^{9-}$ anion. Such product type wavepacket is placed either on the electronic state \textit{A$_{1g}$} or on the\textit{ T$_{1u}$} state(s) of TiO$_6^{8-}$ and for each case, four state dynamical calculation has been performed. The time - dependent wavefunction is employed to calculate the corresponding autocorrelation functions at different time (see Eq. \ref{Eq:Auto-Correlation}), which on Fourier transformation (see Eq. \ref{Eq:Fourier}) generates distinct spectral profiles for each state. Furthermore, the convolution of these individual state spectra yields the comprehensive spectral envelope of the TiO$_6^{8-}$ complex. Though the comparison is made between the calculated PE spectra for the \textit{A$_{1g}$} and \textit{T$_{1u}$} states of TiO$_6^{8-}$ unit and the experimentally observed photoemission~\cite{R_Cord_1985} peaks for the BaTiO$_3$ complex, this is an approximation, but the other theoretical~\cite{Pertosa_Theory_Spectra} spectra for TiO$_6^{8-}$ unit is appropriate. Since the ionization effect is naturally included in the experimental profile, this  comparision  with theoretical one is only within approximation. Figure \ref{fig:sp} demonstrates the peak positions of the photoemission spectra of an isolated TiO$_6^{8-}$ anion in comparison with experimental~\cite{R_Cord_1985} spectra of BaTiO$_3$ lattice and other theoretical~\cite{Pertosa_Theory_Spectra} spectra of TiO$_6^{8-}$ unit, where the present diabatic surface matrix is derived from MRCI(SD)-based APESs as well as CASSCF-based DDR calculated NACTs for TiO$_6^{8-}$ unit. The verticle dotted lines also depict the agreement of the peaks at 1.05 and 3.80 eV. Also note that the difference between the two peak positions in the spectrum ($\sim$ 2.75 eV) closely matches with the calculated global stabilization energy of 2.582 eV, which is attributed to the PJT interactions. At this point, it is important to note that since the TiO$_6^{8-}$ is a unit cell of an infinite crystal, BaTiO$_3$, the ``actual" spectra will be broadened enough due to the dissipation of energy through neighbouring units. Thereby, it is necessary to introduce appropriate value of $\boldsymbol{\tau}$ in the autocorrelation functions (see Eqs. \ref{Eq:damping} and \ref{Eq:cos-function}) so that the damping of energy is quite fast and efficient leading to the PE spectra. This study yields valuable insights for the analysis and interpretation of the observed phenomena, contributing to the wider understanding of the intricate electronic and vibrational interactions in transition metal oxides, but brings little understanding on the nature of cooperative effect originating from the crystal.

\newpage

\subsection{Ferroelectricity} \label{Ferroelectricity}

\noindent
In ferroelectric materials, the spontaneous change of polarization occurs due to the displacement of positive and negative charges within each TiO$_6^{8-}$ unit of the crystal lattice when the material undergoes a phase transition below its critical temperature known as the Curie temperature (\textit{T$_C$}).\cite{lines2001principles} Below \textit{T$_C$}, due to the structural change of unit cell, the crystal structure becomes non-centrosymmetric allowing the formation of permanent electric dipole(s). As the temperature increases approaching to \textit{T$_C$}, the structural distortion diminishes and undergoes transition from a tetragonal to fully cubic phase, where such transition is accompanied by a second order phase change.

\vspace{0.2cm}

\begin{figure}[!htp]
	\centering
	\includegraphics[width=0.4\textwidth]{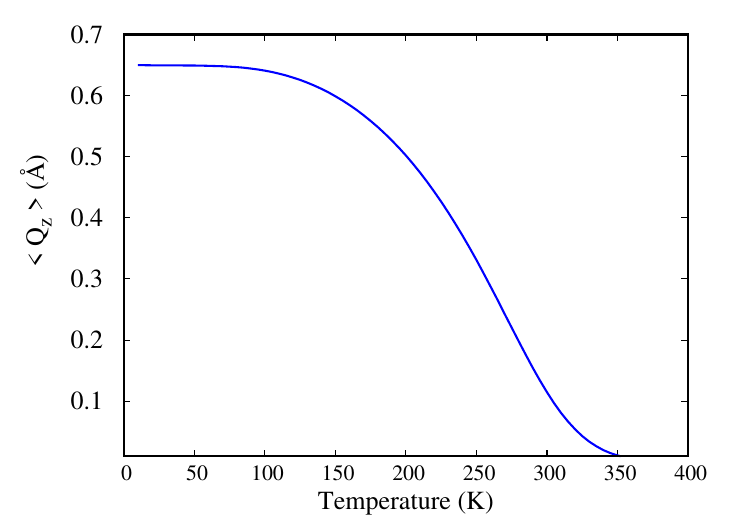}
	\caption{The order Parameter, $\langle$Q$_z\rangle$ in the tetragonal phase exhibit gradual decline, culminating in the vanishing off-centre displacement of titanium ion ($Ti^{4+}$) at \textit{T}= 350 K, indicative of second-order phase transition.}
	\label{fig:Ferro}
\end{figure}

\noindent
 The numerical solution of Eq.~\ref{Eq:Partition_function1} employing newly calculated adiabatic PES (see Section~\ref{ssec:pair2}) provides insights into the system's behaviour on ferroelectricity as shown in Figure \ref{fig:Ferro}. We find that the theoretical calculated Curie temperature (\textit{T$_C$}$\simeq 350$ K) of TiO$_6^{8-}$ unit of barium titanate in Figure \ref{fig:Ferro} closely matches with other theoretical~\cite{Chen_spont_2005,Polinger_2013} and experimentally~\cite{Walter_spont_1949} results for the same crystal (see Figures S9, S10 and S11 in Section S12 of the ESI).
 
 \vspace{0.2cm}
 
 \noindent
  It appears that the PJT effect is the origin of polar instability within a single unit TiO$_6^{8-}$ of the perovskite crystal. This instability leads to an off-centre displacement of the transition metal ion ($Ti^{4+}$), accompanied by significant polarization of valence-shell electronic states and a counter-phase displacement of neighbouring oxygen atoms. The polar distortion is triggered by vibronic PJT coupling between the ground and the excited electronic states of opposite parity, strengthening the covalent bonds between the transition metal and nearby oxygen atoms.

\section{Conclusion} \label{Conclusion}

\noindent
We utilize the coupled cluster method (CCSD/ANO-R1) as implemented in the MOLPRO quantum chemistry package to compute the frequencies of the triply degenerate $t_{2g}$ and $t_{1u}$ vibrational modes of an octahedral TiO$_6^{8-}$ unit  of a BaTiO$_3$ crystal. With these frequencies and displacement vectors of the normal modes for the TiO$_6^{8-}$ unit of BaTiO$_3$ crystal, \textit{ab initio} based adiabatic PESs (MRCI(SD)) and NACTS (DDR) are calculated to explore: (a) how the triply degenerate  $t_{2g}$ vibrational modes lead to JT distortion of the $T_{1u}$ symmetric excited electronic states and (b) at what extent PJT interactions are observed between $A_{1g}$ and $T_{1u}$ states along $t_{1u}$ vibrational modes. In our calculation, it appears that the lowest sheet of $T_{1u}$ state is stabilized only by 0.0182 eV due to JT distortion, whereas the PJT interactions among $A_{1g}$ and $T_{1u}$ states stabilize the $A_{1g}$ state of the order $\sim$ 2.582 eV at  $Q_{t_{1ux}}$ = $Q_{t_{1uy}}$ = $Q_{t_{1uz}}$ = $\pm $0.51  \AA\ with eight such global minima over the nuclear CS.

\vspace{0.2 cm}

\noindent
The couplings (NACTs) among the electronic states ($A_{1g}$ and $T_{1u}$) along the normal modes ($t_{2g}$ and $t_{1u}$)   are evaluated by CASSCF based DDR methodology. While solving the ADT equation (Eqs. \ref{eq:adteq} and \ref{eq:ADTgen}) using \textit{ab initio} calculated NACTs, it is observed that the mixing angles ($\Theta_{23}$ or $\Theta_{34}$) reach $\pi$ (see Figure \ref{fig:t1g_an}) at the completion of a closed contour ($\phi = 2\pi$), indicating the presence of enclosed JT  CI. Conversely, the mixing angles ($\Theta_{12}$ or $\Theta_{13}$ or $\Theta_{14}$) attain zero (see Figure \ref{fig:t1u_an}) at the end of a closed contour ($\phi = 2\pi$), indicative of encapsulated PJT interactions. Furthermore, the presence of such CIs is confirmed by the sign inversion of the diagonal elements of the ADT matrices [A$_{11}$ (=A$_{22}$)]. Finally, these ADT matrices facilitate the generation of smooth, single-valued, and continuous diabatic PESs and couplings.

\vspace{0.2cm}

\noindent
 We construct diabatic PESs and couplings over six (6) pairs of normal modes and carried out TDDVR dynamics with optimized time-dependent basis sets to generate photoemission (PE) spectral profiles for combined four electronic states ($A_{1g}$ and $T_{1u}$). The calculated spectra show reasonably good agreement with experimental profile~\cite{R_Cord_1985}, even though the former one is obtained from single TiO$_6^{8-}$ unit of BaTiO$_3$ crystal lattice and the later is measured for a single crystal of BaTiO$_3$. Though the PJT effect introduces polar instability within a single unit TiO$_6^{8-}$, the cooperative effect of the crystal could be due to higher order polarisation arising from the environment of the lattice. Finally, we also calculate spontaneous polarization leading to ferroelectric properties in BaTiO$_3$ below the Curie temperature using adiabatic PESs, revealing the pseudo Jahn-Teller effect as the origin of polar instability and off-centre displacement of transition metal ions.  Our investigation underscores the effectiveness of first principles-based BBO theory coupled with TDDVR methodology in elucidating the effect of nonadiabaticity on spectral profiles and the use of mean-field Hamiltonian (see Eqs. \ref{Eq:Partition_function} and \ref{Eq:Partition_function1}) for ferroelectric properties of TiO$_6^{8-}$ unit of BaTiO$_3$ crystal lattice.

\section*{Electronic Supplementary Information} 
\begin{itemize}
	\item A detailed theoretical description of TiO$_6^{8-}$ within BaTiO$_3$ is presented in the Electronic Supplementary Information (ESI). This includes figures showing Adiabatic Potential Energy Curves (PECs) and Non-Adiabatic Coupling Terms (NACTs) as well as path integrals and the convergence of photoemission (PE) spectra. Moreover, both experimental and theoretical curves depicting spontaneous polarization variation with temperature are provided. Additionally, an overview of the Time-Dependent Discrete Variable Representation (TDDVR) Formalism is included in the ESI.

\end{itemize}

\section*{Acknowledgements}

M.K.S acknowledges NBCFDC, Govt. of India, Ministry of Social Justice and Empowerment for research fellowship. S.M thanks IACS for research funding and MANIT, Bhopal for the seed grant. S.R is thankful to SERB through project no. SRG/2023/001624. S.A thanks SERB through project no. CRG/2023/000611  and IACS for the research funding and IACS for CRAY supercomputing facility.

\newpage
%

\clearpage

\appendix

\renewcommand{\thesection}{S\arabic{section}}
\renewcommand{\thesubsection}{S\arabic{section}.\arabic{subsection}}
\renewcommand{\thefigure}{S\arabic{figure}}
\renewcommand{\thetable}{S\arabic{table}}
\renewcommand{\thepage}{S\arabic{page}}
\renewcommand{\theequation}{S\arabic{equation}}

\setcounter{section}{0}
\setcounter{figure}{0}
\setcounter{table}{0}
\setcounter{equation}{0}
\setcounter{page}{1}

\begin{center}
	{\LARGE \textbf{Electronic Supplementary Information}} 
\end{center}

\section{Conditions for the JT and PJT Interactions}

\noindent
While exploring the conditions to understand the properties related to the distortions of TiO$_6^{8-}$ unit of BaTiO$_3$ crystal, it appears that the vibrational modes ($t_{2g}$ and $t_{1u}$ ) are responsible for Jahn-Teller (JT) and pseudo Jahn-Teller (PJT) interactions. On the other hand, the $Ti^{4+}$ ion of TiO$_6^{8-}$ unit having a \( d^0 \) electronic configuration leads to a non-degenerate ground state (\( A_{1g} \)), whereas the first excited state is triply degenerate ($ T_{1u} $). The JT and PJT effects originate due to the coupling of electronic state with vibrational modes of appropriate symmetry. In other words, the direct product of irreducible representations for the electronic state and the vibrational mode contains the totally symmetric representation ($ A_{1g} $). The JT distortion  takes place between $^1T_{1u}$ states via $t_{2g}$ normal modes:

\[
A_{1g} \in t_{2g} \otimes [(T_{1u})^2] \equiv g\otimes u \otimes u,
\]

\noindent
whereas the PJT effect involves the coupling between the non-degenerate ground state ($ A_{1g} $) and an excited state ($ T_{1u} $) via vibrational modes ($ t_{1u} $):
\[
A_{1g} \in t_{1u} \otimes [A_{1g} \otimes T_{1u}] \equiv u \otimes g \otimes u.
\]

\noindent
Though our calculations reveal that the JT distortion due to interaction of $ t_{2g} $ normal modes with $T_{1u}$ electronic states is negligibly small, the PJT stabilization for the coupling of $ t_{1u} $ modes with $ A_{1g} $ and $^1T_{1u}$ electronic states is significant. Moreover, \textit{ab initio} calculation along other $u$-symmetric normal modes ($t_{2u}$ and other set of $t_{1u}$) yield approximately the same stabilization energy (-0.65 eV) and thereby, one representative  set of $u$-symmetric normal modes is chosen for the present calculation. This selection simplifies the calculation as predicted before~\cite{Bersuker_2015} to depict the effect of PJT stabilization on photoemission spectra and ferroelectric properties.

\vspace{0.2 cm}

\section{Adiabatic to Diabatic Transformation Equation: Curl Condition}

\noindent
In adiabatic Representation, the kinetically coupled SE can be written as:

\begin{eqnarray}
	- \dfrac{\hbar^2}{2} \left( \vec{\nabla}_R + \vec{\tau} \right)^2  {\psi^{ad}} + (U - E) {\psi^{ad}} = 0.
	\label{eq:adia5}
\end{eqnarray}

\noindent
Using the following transformation:
\begin{eqnarray}
	\psi^{ad} = A \psi^{d},
\end{eqnarray}

\noindent
with $\psi^{ad}$ and  $\psi^{d}$ as adiabatic and  diabatic nuclear wavefunctions, respectively and `$A$' being the adiabatic-to-diabatic transformation matrix, Eq. \ref{eq:adia5} turns into: 

\begin{eqnarray}
	- \dfrac{\hbar^2}{2} \left( \vec{\nabla}_R + \vec{\tau} \right)^2 A {\psi^d} + (U - E) A{\psi^d} = 0,
	\label{eq:diab1}
\end{eqnarray}

\noindent
which appears as:

\begin{eqnarray}
	- \dfrac{\hbar^2}{2}\Big[A\nabla^2_R \psi^d + 2 (\vec{\nabla}_R A + \vec{\tau} A ) \cdot \vec{\nabla}_R \psi^d&+&\left\{(\vec{\tau} + \vec{\nabla}_R) \cdot (\vec{\nabla}_R A + \vec{\tau} A)\right\} \psi^d \Big] A {\psi^d}\nonumber\\
	&+& (U - E) A{\psi^d} = 0.
	\label{eq:adia8}
\end{eqnarray}

If the following constrain is imposed, 

\begin{eqnarray}
	\vec{\nabla}_R A + \vec{\tau} A = 0,
	\label{eq:adt_cond}
\end{eqnarray}

Eq.~\ref{eq:adia8} reduces to the form as given below:

\begin{eqnarray}
	- \dfrac{\hbar^2}{2} A \nabla^2_R \psi^d + (U - E) A \psi^d = 0.
	\label{eq:diab2}
\end{eqnarray}

\noindent
On the other hand, when Eq. \ref{eq:adt_cond} is left multiplied by $A^\dagger$ and  the daggered ($\dagger$) of Eq. \ref{eq:adt_cond} is right multiplied by $A$, we obtain:

\begin{subequations}
	\begin{eqnarray}
		A^{\dagger} \vec\nabla_R A + A^\dagger \vec{\tau} A = 0,\label{eq:adtm1} \\
		(\vec \nabla_R A^\dagger) A - A^\dagger \vec{\tau} A = 0, 
		\label{eq:adtm2}
	\end{eqnarray}
\end{subequations}

\noindent
and then, on adding Eqs. \ref{eq:adtm1} and \ref{eq:adtm2}, we obtain:

\begin{eqnarray}
	A^\dagger \vec \nabla_R A + (\vec \nabla_R A^\dagger) A = 0,\nonumber \\
	\Rightarrow \vec \nabla_R (A^\dagger A) = 0 \nonumber, \\
	\Rightarrow A^\dagger A = \text{const,}
\end{eqnarray}

\noindent
which defines $A$ is an orthogonal matrix.

\vspace{0.2 cm}

\noindent
Therefore, we left multiply Eq. \ref{eq:diab2} by $ A^\dagger $ and obtain the  diabatic representation of SE:

\begin{eqnarray}
	- \dfrac{\hbar^2}{2}\nabla_R^2 \psi^d + (W - E) \psi^d = 0,
	\label{eq:diab}
\end{eqnarray}

where

\begin{eqnarray}
	W = A^\dagger U A.
\end{eqnarray}

\noindent
The couplings between electronic states are incorporated within the potential energy matrix ($W$)~\cite{RBD}.\\

\noindent
While performing BBO-based diabatization of adiabatic PESs and NACTs, it is essential to verify whether the number of electronic states (\(N\)) within the relevant domain of nuclear space can be considered a ``true" SHS for required  numerical accuracy. The existence of such SHS can be assessed by evaluating the matrix elements of Curl Condition~\cite{mantu_curl} associated with the NACTs as outlined below. For any pair of nuclear coordinates (\(p\) and \(q\)), the scalar form of Eq. \ref{eq:adt_cond} can be expressed as:

\begin{equation}
	\nabla_pA + \tau_pA = 0, 
	\label{adt_cond1}
\end{equation}

\begin{equation}
	\nabla_qA + \tau_qA = 0, 
	\label{adt_cond2}
\end{equation}

\noindent
and by taking cross-derivatives to Eqs. \ref{adt_cond1} and \ref{adt_cond2}, we obtain:
\begin{subequations}
	\begin{equation}
		\nabla_q\nabla_pA + \Big(\dfrac{\partial}{\partial q}\tau_p\Big)A+ \tau_p \dfrac{\partial}{\partial q}A = 0, 
		\label{adt_cond3}
	\end{equation}
	
	\begin{equation}
		\nabla_p\nabla_qA + \Big(\dfrac{\partial}{\partial p}\tau_q\Big)A+ \tau_q \dfrac{\partial}{\partial p}A = 0. 
		\label{adt_cond4}
	\end{equation}
\end{subequations}
\noindent
Since the matrix element of \( A \) are analytic functions of the nuclear coordinates (\( p \) and \( q \)) and $\nabla_q\nabla_pA=\nabla_p\nabla_qA$, we obtain the following Curl Condition:

\begin{equation}
	\dfrac{\partial}{\partial p}\tau^{ij}_q-\dfrac{\partial}{\partial q}\tau^{ij}_p=(\tau_q\tau_p)_{ij}-
	(\tau_p\tau_q)_{ij},
	\label{eq:curl}
\end{equation}

\noindent
where the matrix elements (analogous to Yang-Mills field) over $p-q$ plane are represented as:

\begin{equation}
	F_{pq}^{ij}=\Big[\dfrac{\partial}{\partial p}\tau^{ij}_q-\dfrac{\partial}{\partial q}\tau^{ij}_p
	\Big]-\Big[(\tau_q\tau_p)_{ij}-
	(\tau_p\tau_q)_{ij}\Big]=Z_{pq}^{ij}-C_{pq}^{ij}.
	\label{eq:curl1}
\end{equation}

\noindent
When the sub-space is complete, the magnitude of \( F_{pq}^{ij} \) should be zero (0) over the relevant domain of nuclear configuration space, ensuring that the non-removable component of the NACTs becomes negligibly small. If this condition is not satisfy for the chosen SHS, it becomes necessary to enlarge the sub-space in order to diabatize the SE in a ``true" sense, achieving the desired level of accuracy.\\

\section{Explicit Expression of ADT Equations and NACTs for Four state ($N=4$) sub-Hilbert space~\citep{sarkar_jpca112}}

\subsection{ADT Equations}

\begin{footnotesize}
	\begin{subequations}
		\begin{eqnarray*}
			\boldsymbol{\nabla}_R\Theta^{12}&=& -\boldsymbol{\tau}^{12}-\sin\Theta^{12}\tan\Theta^{13}\boldsymbol{\tau}^{13}-\cos\Theta^{12}\tan\Theta^{13}
			\boldsymbol{\tau}^{23}-\sin\Theta^{12}\sec\Theta^{13}\tan\Theta^{14}\boldsymbol{\tau}^{14}-\cos\Theta^{12}\sec\Theta^{13}\tan\Theta^{14}
			\boldsymbol{\tau}^{24} \\
			\boldsymbol{\nabla}_R\Theta^{13}&=& -\cos\Theta^{12}\boldsymbol{\tau}^{13}+\sin\Theta^{12}\boldsymbol{\tau}^{23}-\cos\Theta^{12}\sin\Theta^{13}
			\tan\Theta^{14}\boldsymbol{\tau}^{14}+\sin\Theta^{12}\sin\Theta^{13}\tan\Theta^{14}\boldsymbol{\tau}^{24}-\cos\Theta^{13}\tan\Theta^{14}\boldsymbol{\tau}^{34}\\
			\boldsymbol{\nabla}_R\Theta^{23}&=&-\cos\Theta^{13}[\boldsymbol{\tau}^{13}\sin\Theta^{12}\sec^2\Theta^{13}+\cos\Theta^{23}\sec\Theta^{14}(\boldsymbol{\tau}^{34}-\boldsymbol{\tau}^{24}\sin\Theta^{12}\tan\Theta^{13})\tan\Theta^{24}\nonumber\\
			&&+\sin\Theta^{12}\sec\Theta^{13}\tan\Theta^{13}\boldsymbol{\tau}^{14}\tan\Theta^{14}+\sin\Theta^{23}\boldsymbol{\tau}^{14}\sec\Theta^{14}\tan\Theta^{24}+\cos\Theta^{12}\lbrace\boldsymbol{\tau}^{23}\sec^2\Theta^{13}\nonumber \\
			&&+\tan\Theta^{13}\cos\Theta^{23}\boldsymbol{\tau}^{14}\sec\Theta^{14}
			\tan\Theta^{24}+\sec\Theta^{13}(\tan\Theta^{13}\boldsymbol{\tau}^{24}
			\tan\Theta^{14}\nonumber \\
			&&+\sin\Theta^{23}\boldsymbol{\tau}^{24}\sec\Theta^{14}
			\tan\Theta^{24})\rbrace]\\
			\boldsymbol{\nabla}_R\Theta^{14}&=&-\cos\Theta^{12}\cos\Theta^{13}\boldsymbol{\tau}^{14}+\sin\Theta^{12}\cos\Theta^{13}\boldsymbol{\tau}^{24}
			+\sin\Theta^{13}\boldsymbol{\tau}^{34}\\
			\boldsymbol{\nabla}_R\Theta^{24}&=&\sin\Theta^{23}(-\sin\Theta^{12}\sin\Theta^{13}\boldsymbol{\tau}^{24}\sec\Theta^{14}+
			\cos\Theta^{13}\boldsymbol{\tau}^{34}\sec\Theta^{14})-
			\sin\Theta^{12}\cos\Theta^{23}\boldsymbol{\tau}^{14}\sec\Theta^{14}\nonumber \\
			&&+\cos\Theta^{12}(\sin\Theta^{13}
			\sin\Theta^{23}\boldsymbol{\tau}^{14}\sec\Theta^{14}-\cos\Theta^{23}\boldsymbol{\tau}^{24}\sec\Theta^{14}) \\
			\boldsymbol{\nabla}_R\Theta^{34}&=&\sin\Theta^{12}\lbrace-\sin\Theta^{23}\boldsymbol{\tau}^{14}\sec\Theta^{14}\sec\Theta^{24}+\sin\Theta^{13}\cos\Theta^{23}\boldsymbol{\tau}^{24}\sec\Theta^{14}\sec\Theta^{24}\rbrace+\cos\Theta^{12}[-\sec\Theta^{24}\nonumber\\
			&&\lbrace\sin\Theta^{13}\cos\Theta^{23}\boldsymbol{\tau}^{14}\sec\Theta^{14}+\sin\Theta^{23}
			\boldsymbol{\tau}^{24}\sec\Theta^{14}\rbrace]-\cos\Theta^{13}\cos\Theta^{23}\boldsymbol{\tau}^{34}\sec\Theta^{14}\sec\Theta^{24}
		\end{eqnarray*}
		\label{eq:adteq1}
	\end{subequations}
\end{footnotesize}

\subsection{NACTs}
\begin{footnotesize}
	\begin{subequations}
		\begin{eqnarray*}
			\boldsymbol{\tau^{12}}&=&-\cos\Theta^{13}\cos\Theta^{23}\cos\Theta^{14}\cos\Theta^{24}\boldsymbol{\nabla}_R\Theta^{12}-\sin\Theta^{23}\cos\Theta^{14}\cos\Theta^{24}\boldsymbol{\nabla}_R\Theta^{13}-\sin\Theta^{24}\boldsymbol{\nabla}_R\Theta^{14}\\
			\boldsymbol{\tau^{13}}&=&\cos\Theta^{13}\sin\Theta^{23}\cos\Theta^{14}\cos\Theta^{34}\boldsymbol{\nabla}_R\Theta^{12}+\cos\Theta^{13}\cos\Theta^{23}\cos\Theta^{14}\sin\Theta^{24}\sin\Theta^{34}\boldsymbol{\nabla}_R\Theta^{12}-\cos\Theta^{23}\cos\Theta^{14}\cos\Theta^{34}\boldsymbol{\nabla}_R\Theta^{13}\nonumber\\
			&&+\sin\Theta^{23}\cos\Theta^{14}\sin\Theta^{24}\sin\Theta^{34}\boldsymbol{\nabla}_R\Theta^{13}\\
			\boldsymbol{\tau^{23}}&=&-\sin\Theta^{13}\cos\Theta^{24}\cos\Theta^{34}\boldsymbol{\nabla}_R\Theta^{12}-\cos\Theta^{13}\sin\Theta^{23}\sin\Theta^{14}\sin\Theta^{24}\cos\Theta^{34}\boldsymbol{\nabla}_R\Theta^{12}-\cos\Theta^{13}\cos\Theta^{23}\sin\Theta^{14}\sin\Theta^{34}\boldsymbol{\nabla}_R\Theta^{12}\nonumber\\
			&&+\cos\Theta^{23}\sin\Theta^{14}\sin\Theta^{24}\cos\Theta^{34}\boldsymbol{\nabla}_R\Theta^{13}-\sin\Theta^{23}\sin\Theta^{14}\sin\Theta^{34}\boldsymbol{\nabla}_R\Theta^{13}-\cos\Theta^{24}\cos\Theta^{34}\boldsymbol{\nabla}_R\Theta^{23}-\sin\Theta^{34}\boldsymbol{\nabla}_R\Theta^{24}\\
			\boldsymbol{\tau^{14}}&=&-\cos\Theta^{13}\sin\Theta^{23}\cos\Theta^{14}\sin\Theta^{34}\boldsymbol{\nabla}_R\Theta^{12}+\cos\Theta^{13}\cos\Theta^{23}\cos\Theta^{14}\sin\Theta^{24}\cos\Theta^{34}\boldsymbol{\nabla}_R\Theta^{12}+\cos\Theta^{23}\cos\Theta^{14}\sin\Theta^{34}\boldsymbol{\nabla}_R\Theta^{13}\nonumber\\
			&&+\sin\Theta^{23}\cos\Theta^{14}\sin\Theta^{24}\cos\Theta^{34}\boldsymbol{\nabla}_R\Theta^{13}-\cos\Theta^{24}\cos\Theta^{34}\boldsymbol{\nabla}_R\Theta^{14}\\
			\boldsymbol{\tau^{24}}&=&\sin\Theta^{13}\cos\Theta^{24}\sin\Theta^{34}\boldsymbol{\nabla}_R\Theta^{12}+\cos\Theta^{13}\sin\Theta^{23}\sin\Theta^{14}\sin\Theta^{24}\sin\Theta^{34}\boldsymbol{\nabla}_R\Theta^{12}-\cos\Theta^{13}\cos\Theta^{23}\sin\Theta^{14}\cos\Theta^{34}\boldsymbol{\nabla}_R\Theta^{12}\nonumber\\
			&&-\cos\Theta^{23}\sin\Theta^{14}\sin\Theta^{24}\sin\Theta^{34}\boldsymbol{\nabla}_R\Theta^{13}-\sin\Theta^{23}\sin\Theta^{14}\cos\Theta^{34}\boldsymbol{\nabla}_R\Theta^{13}+\cos\Theta^{24}\sin\Theta^{34}\boldsymbol{\nabla}_R\Theta^{23}-\cos\Theta^{34}\boldsymbol{\nabla}_R\Theta^{24}\\
			\boldsymbol{\tau^{34}}&=&-\sin\Theta^{13}\sin\Theta^{24}\boldsymbol{\nabla}_R\Theta^{12}+\cos\Theta^{13}\sin\Theta^{23}\sin\Theta^{14}\cos\Theta^{24}
			\boldsymbol{\nabla}_R\Theta^{12}-\cos\Theta^{23}\sin\Theta^{14}\cos\Theta^{24}\boldsymbol{\nabla}_R\Theta^{13}\nonumber\\
			&&-\sin\Theta^{24}\boldsymbol{\nabla}_R\Theta^{23}-\boldsymbol{\nabla}_R\Theta^{34}
		\end{eqnarray*}   
	\end{subequations}
\end{footnotesize}

\section{Adiabatic Potential Energy Curves (PECs) and NACTs for TiO$_6^{8-}$ unit of BaTiO$_3$ crystal} 

\begin{figure}[h]
	\centering
	\includegraphics[width=0.5\linewidth]{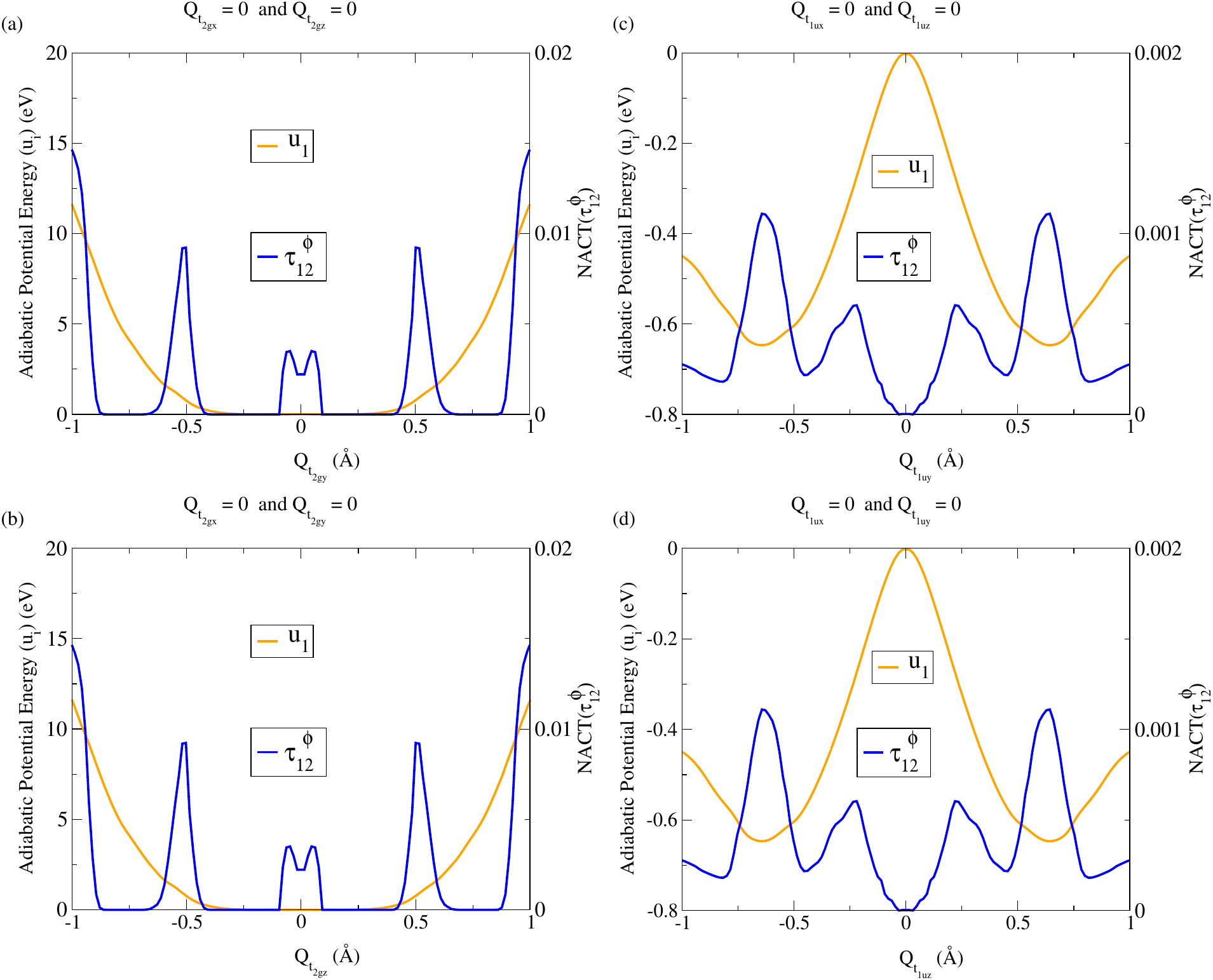}
	\caption{For TiO$_6^{8-}$ unit, 1D curves of lowest adiabatic PES ($u_1$) and the associated NACT ($\tau_{12}^{\phi}$) are presented along (a) $Q_{t_{2gy}}$, (b) $Q_{t_{2gz}}$, (c) $Q_{t_{1uy}}$, and (d) $Q_{t_{1uz}}$ normal modes keeping the other normal modes fixed at zero (0).}
	\label{fig:Norm_Prob}
\end{figure}

\newpage
\section{Local Topographic Parameters and Adiabatic Potential Energy Surfaces Around the 2-3 Conical Intersection (CI)}
\noindent
We have calculated the local topographic parameter at the geometry of one of the `2-3' CI~\cite{Yarkony_jcp_2001}, where two independent geometrical distortions can linearly break the degeneracy, commonly known as \emph{branching plane vectors}. These two vectors are mainly represented as $\vec{g}$ (the half-difference between the gradients of the two intersecting states) and $\vec{h}$ (the non-adiabatic coupling vectors between the two states). In order to calculate the local topographic parameters (tilt parameters ($s_x$ and $s_y$), pitch of cone ($d_{gh}$) and asymmetry of cone ($\Delta_{gh}$)) for a double cone, we have employed COLUMBUS quantum chemistry package~\cite{Barbatti_Columbus_jpca_2005}, where the calculated parameters are tabulated below: 

\begin{table}[!ht]\label{table:topo}
	\centering
	\begin{tabular}{ |c|c| c| c|} 
		\hline
		$s_x$ (eV/\r{A})   & $s_y$ (eV/\r{A})& $\Delta_{gh}$ & $d_{gh}$ (eV/\r{A})   \\
		\hline
		-0.046 & -0.004 & 0.012 & 0.005 \\
		\hline
	\end{tabular}
\end{table}

\noindent
On the basis of above local topographic parameters, the model double cone adiabatic potential 
energy surfaces (PESs) have been calculated using the following functional form: 
\begin{eqnarray}
	U_2 = s_{x} \cdot x + s_{y} \cdot y - d_{gh} \big [\frac{x^2 + y^2}{2} + \Delta_{gh} \frac{(x^2 - y^2)}{2}\big]^{1/2} \\
	U_3 = s_{x} \cdot x + s_{y} \cdot y + d_{gh} \big [\frac{x^2 + y^2}{2} + \Delta_{gh} \frac{(x^2 - y^2)}{2}\big]^{1/2}
\end{eqnarray}
and the associated PESs are represented in the following figure:

\newpage

\begin{figure}[!htp]
	\centering
	\includegraphics[width=0.7\linewidth]{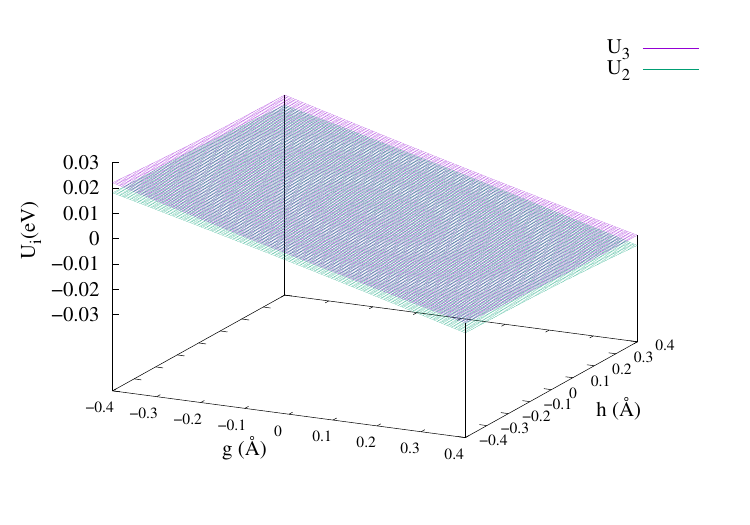}
	\caption{Model adiabatic PESs around `2-3' CI along two branching plane vectors ($g$ and $h$)}
	\label{fig:conical}
\end{figure}

\noindent
Figure~\ref{fig:conical} shows that the calculated model PESs using topographic parameters
vary linearly at the close vicinity of CI. In other words, degeneracy is lifted linearly 
around the `2-3' CI indicating that the intersections are ``conical" not glancing. \\

\begin{figure}[!htp]
	\centering
	\includegraphics[width=0.8\linewidth]{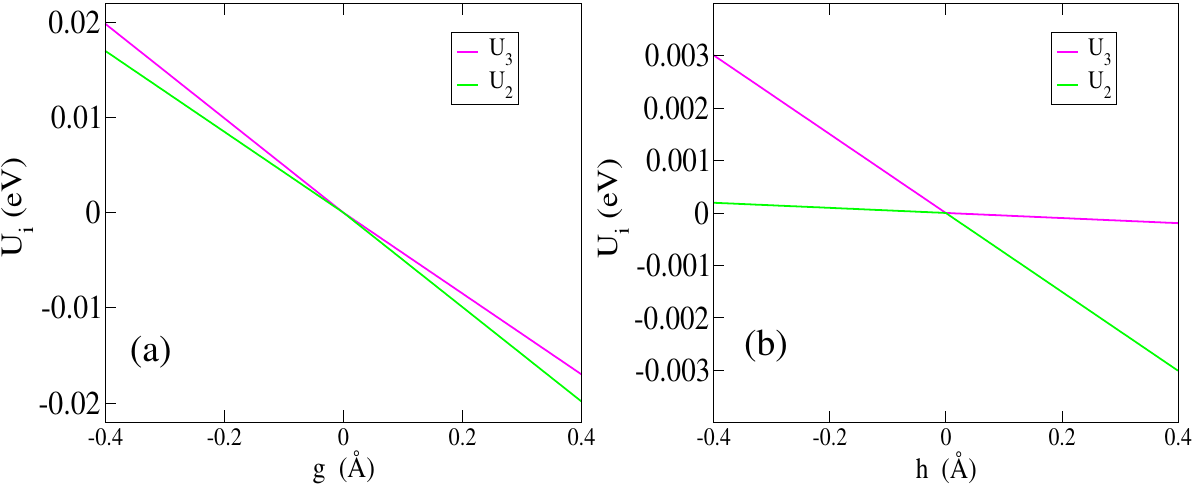}
	\caption{1D curves of model adiabatic PESs around `2-3' CI are presented along branching plane vector (a) g and  (b) h. In each case, the PECs are plotted while keeping the other coordinate fixed at zero: (a)  h = 0  and (b)  g = 0.}
	\label{fig:conical1}
\end{figure}

\section{{\bf The $t_{1u}$ normal mode vibration of TiO$_6^{-8}$ unit}}

\noindent
The triply degenerate vibrational modes ($t_{1u}$) are purely bending modes along \emph{XZ}, \emph{XY} and \emph{YZ} plane. In \textit{ab-initio} calculation, two modes are chosen along $X$ and $Y$ direction to locate the PJT minima.

\begin{figure}[!htp]
	\begin{minipage}[b]{\linewidth}
		\centering
		\includegraphics[width=\textwidth]{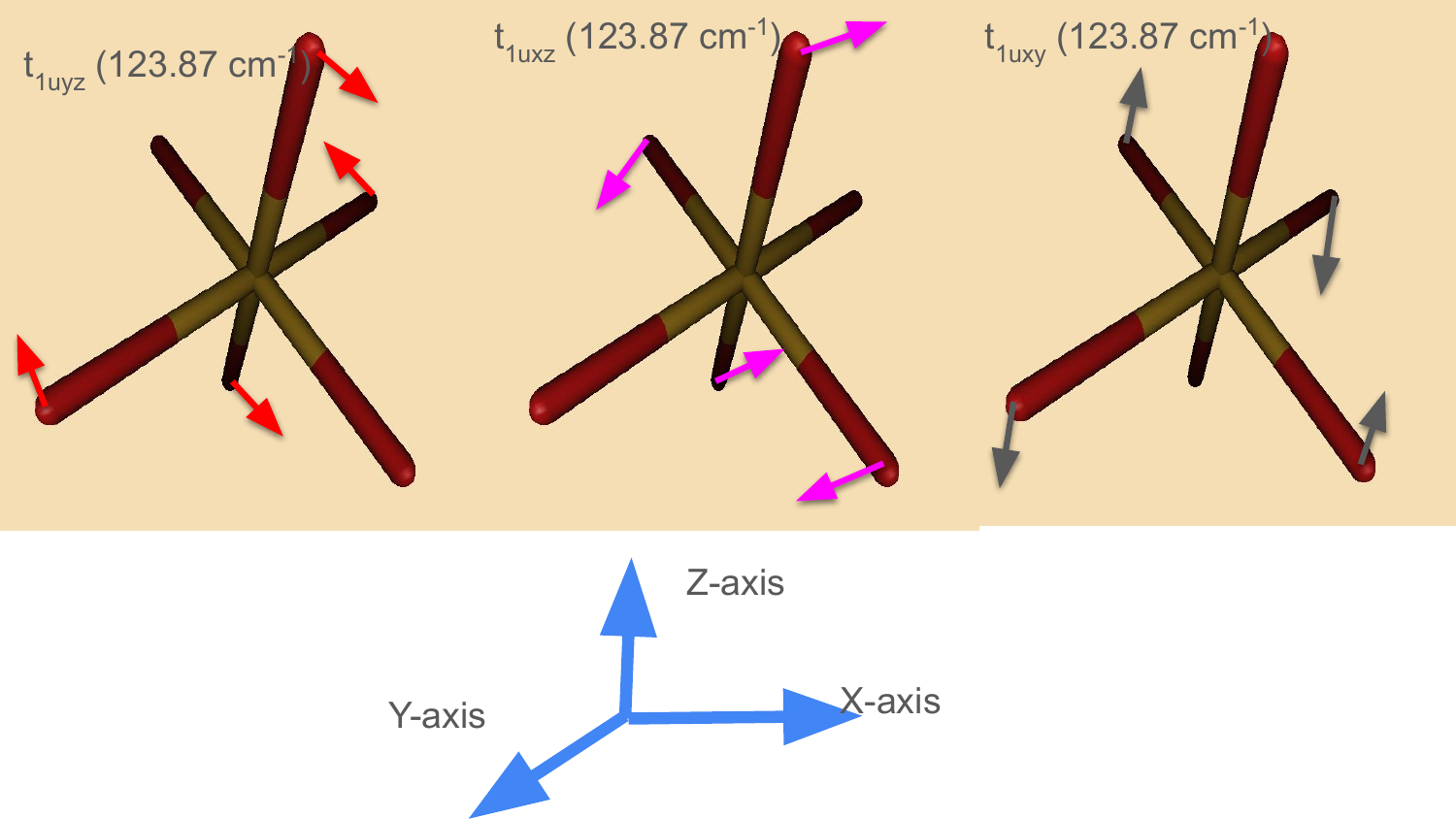}
		\caption{Schemetic picture of $t_{1u}$ modes of TiO$_6^{-8}$ calculated at CCSD level for O$_h$ configuration (Ti-O = 2.0 \r{A}). The calculated frequency is 123.87 cm$^{-1}$. The arrow shows the bending motion of Ti-O bond.} 
	\end{minipage}
\end{figure}

\newpage

\section{{\bf Optimized Geometry of TiO$_6^{-8}$ on the ground state}}

\begin{figure}[!htp]
	\begin{minipage}[b]{\linewidth}
		\centering
		\includegraphics[width=0.9\textwidth]{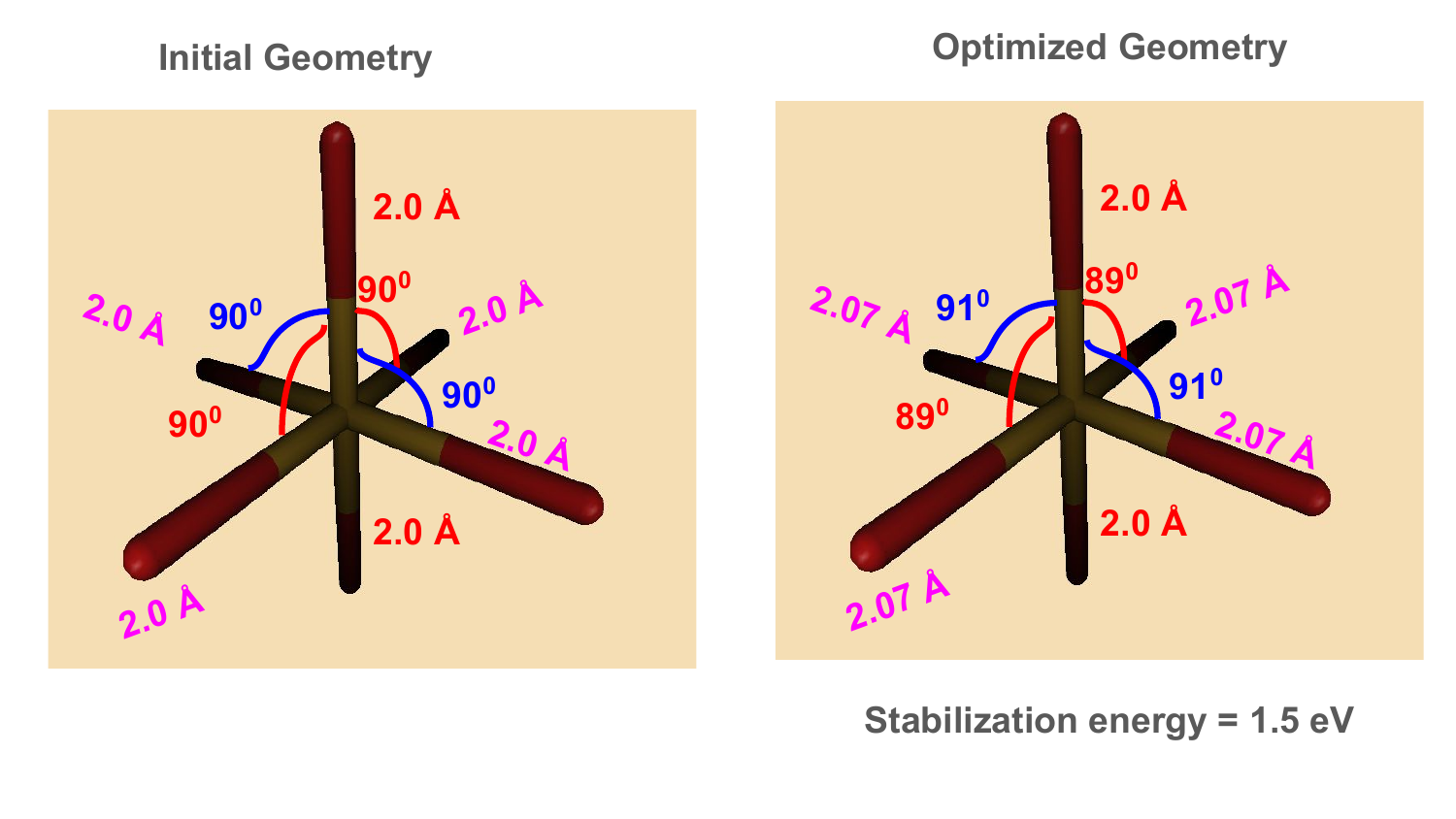}
		\caption{The optimized geometry of TiO$_6^{-8}$ on the ground state. } 
	\end{minipage}
\end{figure}

\section {{\bf Geometry of PJT Stabilized TiO$_6^{-8}$ on the ground state}}

\begin{figure}[!htp]
	\begin{minipage}[b]{\linewidth}
		\centering
		\includegraphics[width=0.5\textwidth]{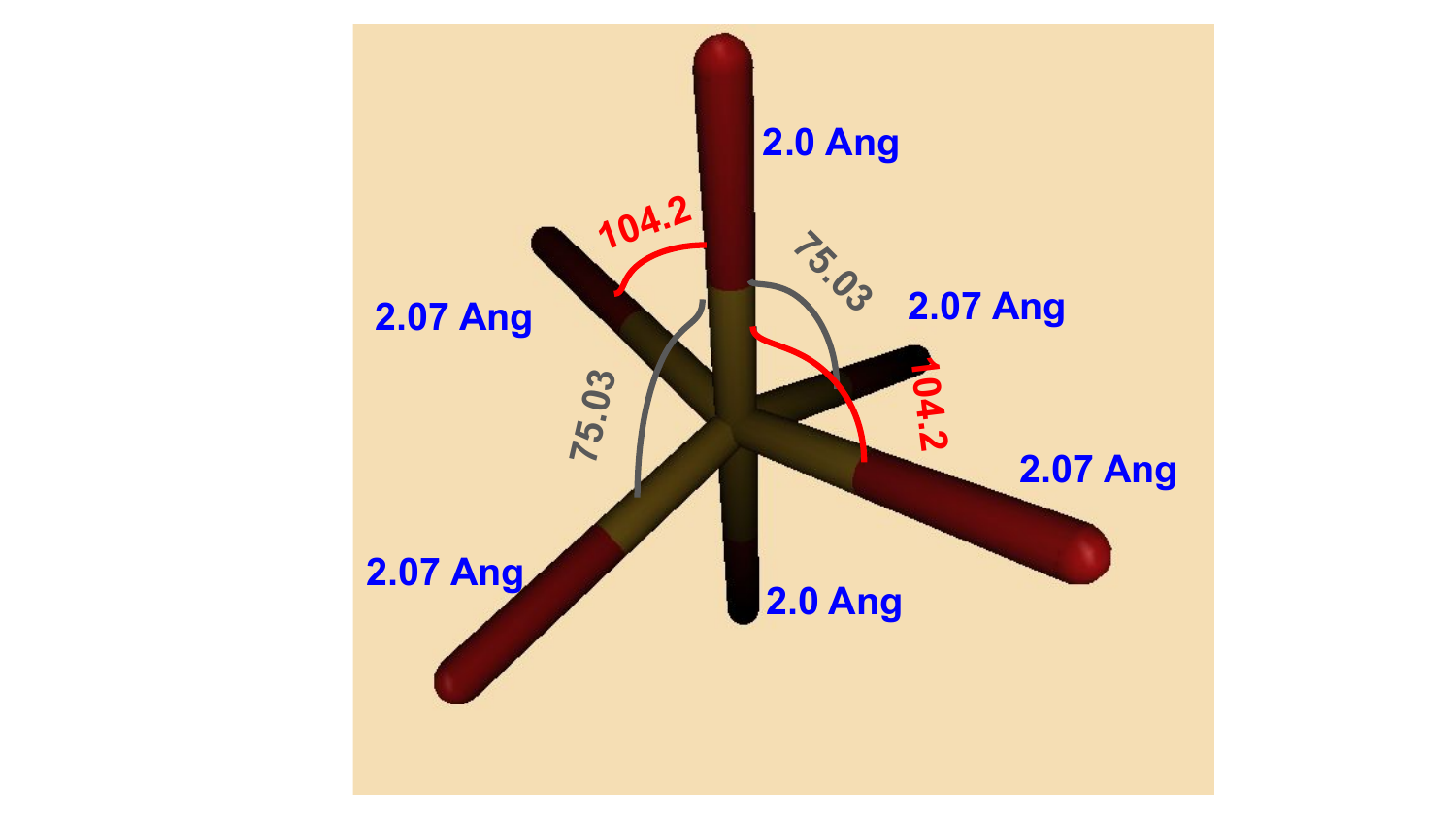}
		\caption{The geometry of PJT stabilized TiO$_6^{-8}$ on the ground state (extracted from \emph{ab-initio calculation} at $\rho = 8.2$ and $\phi$ = 0).} 
	\end{minipage}
\end{figure}

\section{Integration Paths for Stiff Differential Equations~\citep{Mantu_2023}} 

\begin{figure}[!htp]
	\centering
	\includegraphics[width=0.9\linewidth]{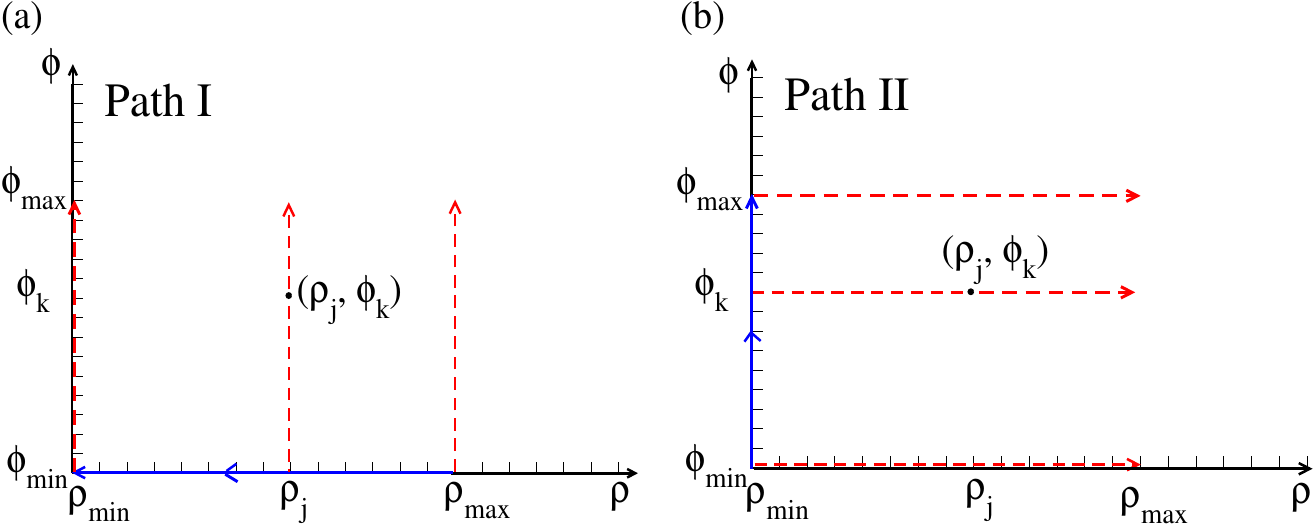}
	\caption{Panels (a) and (b) depict the numerical solution of the stiff differential equations along path I and path II. The integration process involves solving the equations first along the bold (blue) line and subsequently along the dotted (red) lines in each case.}
	\label{fig:Paths}
\end{figure}

\section{An Overview on Time-Dependent Discrete Variable Representation (TDDVR) Formalism}\label{sec:tddvr}

\noindent
The TDDVR formalism has been implemented over wide range of problems involving nuclear dynamics on low-dimensional model~\citep{SSJCS122,
	SSMP107} systems as well as multi-dimensional multi-surface chemical processes~\citep{subhankar_ijqc111,subhankar_jcp130,subhankar_jcs124,SSJPCA118}. While dealing with TDDVR dynamics in multi-dimensional multi-surface molecular systems, we can formulate the time-dependent  Schr\"odinger   Equation (TDSE) within the diabatic framework in the following manner,
\begin{eqnarray}
	i\hbar \frac{\partial}{\partial t}\Xi (\{Q_k\},t) = [{\bf \widehat{T}}_{nuc} {\{Q_k\}} 
	+ {\bf \hat{W}}(\{Q_k\})] \Xi (\{Q_k\},t), \label{eq:tdse2}
\end{eqnarray}
where ${\bf \hat{T}}_{nuc} {\{Q_k\}}$ (= $\hat{T}_{nuc} {\{Q_k\}} \cdot {\bf I}$) and ${\bf \hat{W}}(\{Q_k\})$ denote the
kinetic energy operator and diabatic PES matrix, respectively, let say, expressed in terms of normal mode coordinates, ${\{Q_k\}}$. For systems with $N$ coupled electronic states, the nuclear wavefunction can be represented as follows,
\begin{eqnarray}
	\Xi(\{Q_k\},t)\equiv \left( \begin{array}{c}
		\psi_1 (\{Q_k\},t)\\
		\psi_2 (\{Q_k\},t)\\
		\psi_3 (\{Q_k\},t)\\
		\vdots \\
		\psi_N (\{Q_k\},t)\\
	\end{array}\right) \label{eq:2dwf}
\end{eqnarray}
where $\Xi (\{Q_k\},t)$ is normalized 
[$\int\Xi^{\dagger}(\{Q_k\},t) \Xi (\{Q_k\},t)\prod_{k=1}^p dQ_k$= 1] at any time $t$.\\

\noindent
In TDDVR formalism, the wavefunction corresponding to the $l^{th}$ PES, denoted as $\psi_l(\{Q_k\},t)$ (see Eq. \ref{eq:2dwf}) is represented using TDDVR basis functions ($\chi_{i_k}(Q_k,t)$) for a total of $p$ normal modes.
\begin{equation}
	\psi_l(\{Q_k\},t)=\sum_{i_1i_2...i_p} c_{i_1i_2....i_p,l}(t) \prod_{k=1}^p \chi_{i_k}(Q_k,t)
	\label{eq:psil2}
\end{equation}
Alternatively, the TDDVR basis functions can be written in terms of Discrete Variable Representation (DVR) basis and time-evolving plane waves:
\begin{eqnarray}
	\chi_{i_k}(Q_k,t)&=& \phi(Q_k,t) \sum_{n=0}^{N_k} \zeta_n^{\ast}(x_{i_k}) \zeta_n(x_k) \label{eq:psiik2d}\\
	&=& \sum_{n=0}^{N_k} \zeta_n^{\ast}(x_{i_k}) \Phi_n(Q_k,t), \label{eq:xiphi}
\end{eqnarray}
where the plane wave takes the following form:	
\begin{eqnarray}
	\phi(Q_k,t) &=& \pi^{1/4} \exp \Big( \frac{i}{\hbar}\big\{ p_{Q^c_k}(t) [Q_k - Q^c_k(t)]\big\}\Big).
\end{eqnarray}
In the DVR basis functions, harmonic oscillator eigenfunctions are chosen as the primitive bases (Eq. \ref{eq:psiik2d}),
\begin{eqnarray}
	\zeta_n (x_k) &=& \Big(\frac{2ImA_k}{\pi\hbar}\Big)^{1/4} \frac{1}{\sqrt{n!2^n \sqrt\pi}} 
	\exp \{-(x_k)^2/2\} H_n(x_k), \label{eq:xin}
\end{eqnarray}
where  
\begin{eqnarray}
	x_k &=& \sqrt{\frac {2ImA_k}{\hbar}}(Q_k - Q_k^c(t)). \label{eq:xk}
\end{eqnarray}
\noindent	
In a similar way, the roots of $N_{k^{th}}$ Hermite polynomial, $H_{N_k}(x_k)$~\citep{SACPL321} attain the following expression,
\begin{eqnarray}
	x_{i_k}&=& \sqrt{\frac {2ImA_k}{\hbar}}(Q_{i_k}(t) - Q^c_k(t)), \label{eq:xiik}
\end{eqnarray}

\noindent
resulting into the following expression of TDDVR grid-points:
\begin{eqnarray}
	Q_{i_k}(t) = Q^c_k(t) + \sqrt{\frac{\hbar}{2 Im A_k}} x_{i_k}.  \label{eq:qik2d}
\end{eqnarray}	

\noindent
It's important to note that while the centre of the wavepacket ($\{Q^c_k\}$) and its momentum ($\{p_{Q^c_k}\}$) are assumed to be time-varying, the imaginary part of the width (${ImA_k}$) is introduced as time-independent.~\citep{SAJCP118} In simpler terms, the time dependency arises from the TDDVR grid points, $Q_{i_k}$s, which are influenced by the variables, $Q^c_k(t)$ and $p_{Q^c_k}(t)$.\\

\noindent
The Gauss-Hermite (G-H) basis functions for the $k^{th}$  normal mode ($\Phi_n(Q_k,t)$ in Eq. \ref{eq:xiphi}) are confirmed to be orthonormal~\citep{souvik_irpc37}, with the ground state representing the Gaussian Wave Packet (GWP). Similarly, the TDDVR basis functions $\chi_{i_k}$s in Eq. (\ref{eq:psiik2d}) for the $k^{th}$ mode adhere to orthogonality, although they do not constitute a normalized set~\citep{souvik_irpc37}.\\

\noindent
By substituting the TDDVR representation of wavefunctions (Eqs. \ref{eq:2dwf} - \ref{eq:xiik}) in the TDSE (Eq. 
\ref{eq:tdse2}), we get the following form of TDDVR matrix equation for the $l^{th}$ PES,
\begin{eqnarray}
	i\hbar {\bf A \dot{C_l}} &=&{\bf H^t_{ll} C_l+ A \sum_{l^{\prime}\neq l} W_{ll^{\prime}} C_{l^{\prime}}} \label{eq:com1}
\end{eqnarray}
which can be transformed into the following convenient (symmetric) form through a similarity transformation,
\begin{eqnarray}
	i\hbar {\bf \dot{D_l}(t) = A^{-1/2} H^t_{ll} A^{-1/2} D_l+\sum_{l^{\prime}\neq l} 
		W_{ll^{\prime}} D_{l^{\prime}}}
\end{eqnarray}
where ${\bf D_l = A^{1/2}C_l}$. The detailed expression of the TDDVR coefficients, $d_{i_1i_2....i_p,l}$ and the specific forms of various component matrices ${\bf \{X^k\}}$ and ${\bf \{Y^k\}}$) of $\bf {H^t}$ are provided in our earlier articles~\citep{souvik_irpc37,soumya_bz,soumya_tfb}. On the other hand, the center of the wavepacket ($\{Q^c_k\}$) and its momentum ($\{p_{Q^c_k}\}$) for the $k^{th}$ mode involve the following classical equations of motion (EOMs):
\begin{eqnarray}
	\dot{Q^c_k}(t) &=&\frac{p_{Q^c_k}(t)}{\mu}, \label{eq:Qceq1}\\
	\dot{p}_{Q^c_k}(t) &=& -\frac{dW(\{Q_k\})}{dQ_k}\Big|_{Q_k(t)=Q_k^c(t)} \,.
	\label{eq:Pceq2}
\end{eqnarray}

\vspace{0.5cm}

\noindent	
While deriving a first principle based explicit expression of $\dot{p}_{Q^c_k}$ for multi-dimensional multi-surface systems, it is necessary to employ the 
Dirac-Frenkel variational principle.~\citep{PAM,souvik_irpc37,soumya_bz,soumya_tfb} Interested readers may refer to the aforementioned works for a thorough explanation of the initialization of wavepackets and their subsequent propagation over the diabatic PESs.

\section{Convergence Test of TDDVR Basis Set for Spectral Profile}

\begin{figure}[!htp]
	\centering
	\includegraphics[width=0.75\textwidth]{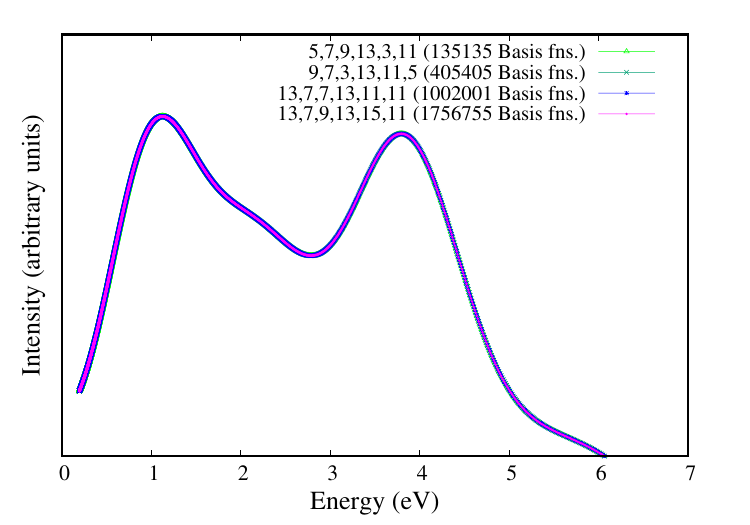}
	\caption{TDDVR calculated spectra obtained from BBO based diabatic PESs are presented for four sets of basis functions. All  sets are almost superimposed with each other and therefore, the set 13,7,7,13,11,11 is used as  the optimized set of basis function.}
	\label{fig:sp1}
\end{figure}

In this present work, while performing the dynamics and calculating the PE spectra of the titanate (TiO$_6^{9-}$) system, we have optimized the basis functions, starting with smaller one, such as (3,3,3,3,3,3) [729 grid points] and then, going to the larger sets, such as (15,15,15,15,15,15) [1,13,90,625 grid points], to find the best combination for accurate calculations in TDDVR dynamics. In the above sets, the sequence of grid points represents six modes ($t_{2g}$ and $t_{1u}$ normal modes) of fundamental vibrational frequencies. On the process of the convergence analysis  to obtain the converged spectral profiles for combined four states ($A_{1g}$ and $T_{1u}$) (13,7,7,13,11,11), total 10,02,001 number of grid points are involved. In order to explore such convergence, four representative PE spectra are shown in the Figure \ref{fig:sp1}.

\newpage

\section{Theoretical and Experimental: Spontaneous Polarization vs Temperature}

\begin{figure}[!htp]
	\centering
	\includegraphics[width=0.6\textwidth]{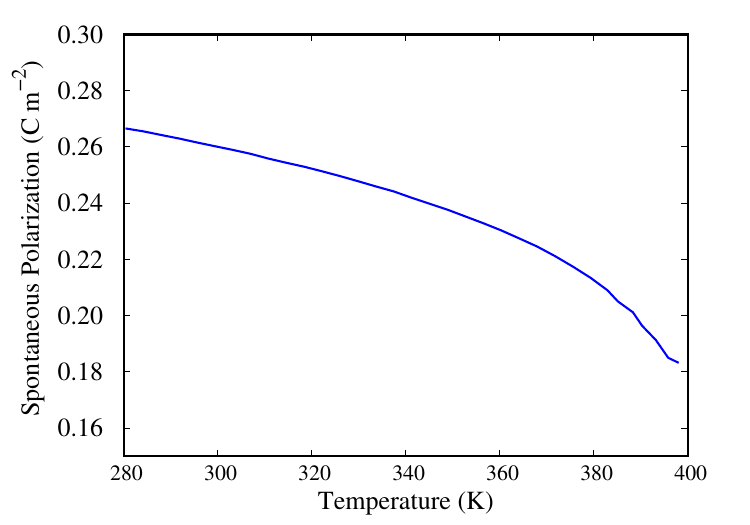}
	\caption{Dependence of polarization on temperature for BaTiO$_3$ single crystal for tetragonal to cubic phase by Li  {\it et al.}~\cite{Chen_spont_2005}.}
	\label{fig:ferro1}
\end{figure}

\begin{figure}[!htp]
	\centering
	\includegraphics[width=0.6\textwidth]{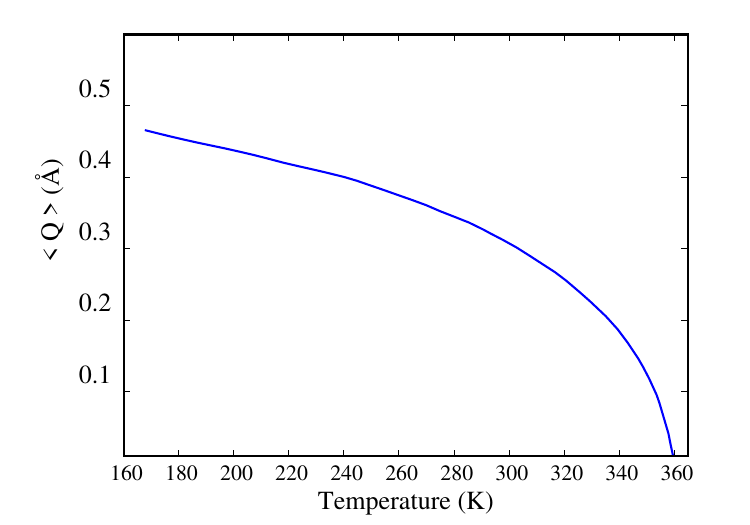}
	\caption{Temperature dependence of the order parameter $\langle Q \rangle$ (in \AA) from tetragonal to cubic phase  by V. Polinger~\cite{Polinger_2013}. }
	\label{fig:ferro2}
\end{figure}

\begin{figure}[!htp]
	\centering
	\includegraphics[width=0.6\textwidth]{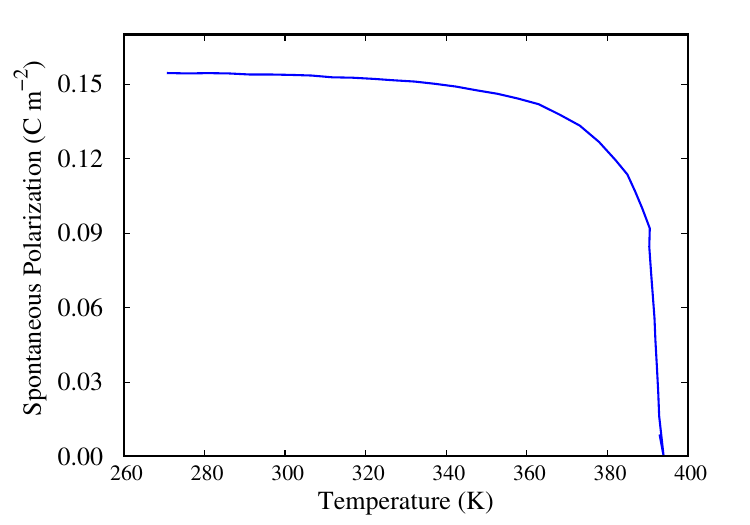}
	\caption{Spontaneous polarization as a function of temperature for tetragonal to cubic phase by Walter J. Merz~\cite{Walter_spont_1949}.}
	\label{fig:ferro3}
\end{figure}

\bibliographystyle{rsc}
\bibliography{ref_cor}

\end{document}